\documentclass[final,3p,times,twocolumn]{elsarticle}

\usepackage{amssymb}

\usepackage{lineno}

\usepackage{url}
\usepackage{epsfig}
\usepackage{epstopdf}
\usepackage[usenames]{xcolor}

\usepackage{tikz}

\newcommand{\reg}{\textsuperscript{\textregistered}}

\journal{Nuclear Instruments and Methods in Physics Research}

\begin{document}

\begin{frontmatter}



\title{The integrated low-level trigger and readout system of the CERN NA62 experiment}


\author[INFNRM2]{R.~Ammendola}
\author[UniPI]{B.~Angelucci\fnref{NowCAEN}}
\author[INFNPI]{M.~Barbanera}
\author[INFNRM]{A.~Biagioni}
\author[Bratislava]{V.~Cerny}
\author[INFNPG]{B.~Checcucci}
\author[INFNPI]{R.~Fantechi}
\author[INFNLNF]{F.~Gonnella\fnref{NowBirmingham}\fnref{ERCBirmingham}}
\author[Bratislava]{M.~Koval\fnref{NowCERN}}
\author[Birmingham]{M.~Krivda}
\author[UniPIeINFNPI]{G.~Lamanna}
\author[UniPG2eINFNPG]{M.~Lupi\fnref{NowCERN}}
\author[INFNRM]{A.~Lonardo}
\author[INFNPG]{A.~Papi}
\author[Birmingham]{C.~Parkinson\fnref{ERCBirmingham}}
\author[UniPIeINFNPI]{E.~Pedreschi}
\author[Louvain]{P.~Petrov}
\author[INFNPI]{R.~Piandani}
\author[UniPIeINFNPI]{J.~Pinzino\fnref{NowCERN}}
\author[INFNPI]{L.~Pontisso\fnref{NowINFNRM}}
\author[INFNLNF]{M.~Raggi}
\author[UniTOeINFNTO]{D.~Soldi}
\author[UniPIeINFNPI]{M.~S.~Sozzi\corref{cor}}
\author[INFNPI]{F.~Spinella}
\author[UniPI]{S.~Venditti\fnref{NowCAEN}}
\author[INFNRM]{P.~Vicini}

\address[INFNRM2]{INFN, Section of Roma Tor Vergata, Via d. Ricerca Scientifica 1, 00133 Roma IT}
\address[UniPI]{Department of Physics, University of Pisa, Largo B. Pontecorvo 3, 56127 Pisa IT}
\address[INFNPI]{INFN, Section of Pisa, Largo B. Pontecorvo 3, 56127 Pisa IT}
\address[INFNRM]{INFN, Section of Roma, Piazzale A. Moro 2, 00185 Roma IT}
\address[Bratislava]{Faculty of Mathematics, Physics and Informatics, Comenius University, Mlynska dolina, 84248 Bratislava, SK}
\address[INFNPG]{INFN, Section of Perugia, Via A. Pascoli 23C, 06123 Perugia IT}
\address[INFNLNF]{INFN, Laboratori Nazionali di Frascati, Via E. Fermi 40, 00044 Frascati IT}
\address[Birmingham]{School of Physics and Astronomy, University of Birmingham, Edgbaston Birmingham B15 2TT UK}
\address[UniPIeINFNPI]{Department of Physics, University of Pisa and INFN, Section of Pisa, Largo B. Pontecorvo 3, 56127 Pisa IT}
\address[UniPG2eINFNPG]{Department of Engineering, University of Perugia and INFN, Section of Perugia, Via A. Pascoli 23C, 06123 
Perugia IT}
\address[Louvain]{Universit\'e Catholique de Louvain, B-1348 Louvain-La-Neuve, BE}
\address[UniTOeINFNTO]{Department of Physics, University of Torino and INFN, Section of Torino, Via P. Giuria 1, 10125 Torino IT}

\cortext[cor]{Corresponding author}
\fntext[NowBirmingham]{Now at School of Physics and Astronomy, University of Birmingham, Edgbaston Birmingham B15 2TT UK.}
\fntext[ERCBirmingham]{Supported by ERC Starting Grant 336581.}
\fntext[NowCAEN]{Now at CAEN S.p.A., Via della Vetraia, 11, 55049 Viareggio LU.}
\fntext[NowCERN]{Now at CERN, CH-1211 Geneva 23, CH.}
\fntext[NowINFNRM]{Now at INFN, Section of Roma, Piazzale A. Moro 2, 00185 Roma IT.}

\begin{abstract}
The integrated low-level trigger and data acquisition (TDAQ) system of the NA62 experiment at CERN is described. 
The requirements of a large and fast data reduction in a high-rate environment for a medium-scale, distributed ensemble of many different 
sub-detectors led to the concept of a fully digital integrated system with good scaling capabilities. 
The NA62 TDAQ system is rather unique in allowing full flexibility on this scale, allowing in principle any information available from the 
detector to be used for triggering.
The design concept, implementation and performances from the first years of running are illustrated.
\end{abstract}


\end{frontmatter}


\section{Introduction}
\label{sec:introduction}

The main goal of the NA62 experiment at CERN \cite{NA62} is the measurement of the branching ratio (BR) of the ultra-rare kaon decay mode
$K^+ \rightarrow \pi^+ \nu \overline{\nu}$: such quantity is predicted in the Standard Model (SM) with a high precision \cite{Theory}, quite 
unusual for hadronic decays, and therefore represents a very powerful probe of possible New Physics. Moreover, in case a discrepancy with 
the SM prediction is measured, the ``theoretically clean'' predictions of such flavour-changing neutral current decay BR would also allow to 
discriminate among different classes of SM extensions.
The downside is that the expected branching ratio is exceedingly small, of order 10$^{-10}$, and the only existing measurement \cite{BNL} is 
based on 7 candidate events, thus lacking any real discriminating power due to its limited precision.
The NA62 experiment, which aims to make a 10\% measurement of the $K^+ \rightarrow \pi^+ \nu \overline{\nu}$ BR using a novel high-energy 
decay-in-flight approach, just concluded its first data taking period.

To achieve the required precision, NA62 must collect $O(10^{13})$ kaon decays, accompanied by a rejection factor of $O(10^{12})$ to suppress 
the huge background from other kaon decays. Part of this suppression must already be made at the trigger level, in order to reduce the amount 
of data that needs to be stored and analysed. Such a large event sample also provides an opportunity to perform many other studies of kaon 
decays, which can result in significant improvements on searches for symmetry violations and in the understanding of QCD and its low-energy 
effective approximation, which are indeed secondary goals of the NA62 experiment. 
Thus the trigger system must be flexible as well as highly selective.
     
The NA62 detector is currently composed of 16 sub-detectors spatially distributed along more than 200~m of beam line, before, around and 
after a 65~m long decay region \cite{NA62}.
Kaons are delivered to the experiment via a high-intensity 75~GeV/$c$ hadron beam, with a beam particle rate close to 1 GHz. 
The decays of beam particles ($\sim$ 6\% being $K^+$) result in an event rate in excess of 10~MHz in the sub-detectors situated after the 
decay region. Excellent time resolution at the trigger level is therefore mandatory, while the minimisation of data collection dead-time, 
and the maximisation of efficiency and reliability, also rank high in importance.
These considerations led to the adoption of a multi-level trigger system.

The lower trigger level, denoted as Level~0 (L0)\footnote{Distinctively reserving positive numbers for software trigger levels.} is implemented 
in hardware and works on data from faster sub-detectors at the full event rate, of order 10~MHz, reducing it by a factor 10 and driving the 
readout of data at 1~MHz to an on-line farm of commercial processors (PC farm), on which further High Level Trigger (HLT) selections are 
performed in software. 
The HLT includes a first level (L1) working on single sub-detector information, reducing the rate to 100~kHz and triggering the completion of 
the readout for the remaining data-heavy sub-detectors (the beam spectrometer and the calorimeters), and a second level (L2) working on the 
full detector information.


This article describes the design and implementation of the L0 trigger level of NA62, which has the distinctive characteristic of being 
fully integrated with the readout system for most sub-detectors.

\section{Overall design}
\label{sec:design}

The L0 trigger system is fully digital, and is designed to work on the main data stream of the experiment: this unification of the (usually 
distinct) data and trigger paths presents several advantages, among which are the reduction of the amount of hardware, and the complete 
control and monitoring of the trigger performance, reproducible at bit level on collected control data. 
Most importantly, the above approach imposes no limitations in principle on the kind of trigger processing which can be performed. 
This is an important asset in an experiment using a novel approach, in which trigger conditions are expected to evolve both because of the 
experience gained during data-taking, and the possibility of expanding the physics programme.

The reason why such approach was not normally implemented by earlier experiments, which usually rely on separated hardware (often partly 
analogue) trigger systems handling a reduced sub-set of detector information, is related to the amount of data which needs to be read-out 
autonomously before a trigger is issued and temporarily stored while the trigger decision is being evaluated. 
Current technology, and in particular the decreasing cost of digital memories, allows a high-rate experiment with a total channel count of 
order 100 thousand such as NA62 to fully store its digitized data for a quite long time compared to the average inverse event rate.

Besides the main concept of full integration of the L0 trigger and data acquisition systems, two more key points in the design were the use of 
a single unified path for trigger and control of individual system boards, and the use of common Gigabit Ethernet (GbE) output data links.
The first point follows an established trend in HEP experiments, and allowed the use of existing hardware developed for LHC experiments.
The second one, besides its advantages in terms of cost and simplicity, resulted in a large flexibility and scalability in the PC farm through
a switched network. This flexibility was exploited to adapt to different running conditions, for example: by changing the bandwidth sharing 
between L0 trigger information and main read out data; and by changing to higher-performance processors in the PC farm without needing to 
change the L0 system.

As mentioned, the L0 trigger system works on the full event rate, in excess of 10~MHz, and its rejection factor of about 10 is expected to 
match the design maximum readout rate of 1~MHz for most sub-detectors.
The latency of the L0 trigger system was chosen to be 1~ms, a rather large value compared to usual implementations, to possibly allow 
the use of massively-parallel processors already at this level, as the recent dramatic increase in performance for such devices suggested 
(see section \ref{sec:gpu}).


In the following we describe the overall scheme of the common integrated readout and L0 trigger system used for most detectors in NA62, whose 
elements are detailed in individual sections, starting with information on the backbone structure which allows the entire system to work 
in a tightly-synchronized way (section \ref{sec:common}). 
As a fixed-target experiment running at the CERN SPS accelerator, NA62 receives beam in a periodic way, with bursts of a few seconds duration 
every several tens of seconds (up to about one minute). The time structure of bursts is (roughly) constant during each data-taking run of a 
few-hours duration, but can change significantly on a daily level, depending on the accelerator working mode. Bursts naturally define coherent 
data-taking units, down to the level of a final data file for permanent storage, identified by a unique burst identifier. This was chosen to be 
the UNIX time of a conveniently chosen instant of the burst itself, encoded as a signed 32-bit number and centrally assigned by the PC farm 
management system and broadcast over the network.
Sub-systems' synchronization is achieved through the use of the common timestamp, defined as a 32-bit unsigned word with 25~ns least significant 
bit. 
The timestamp is locally generated in each individual electronic board from the common distributed 40~MHz clock, and fully synchronized 
throughout the entire system at the start of every burst.

Events are processed by custom electronics and stored into temporary buffers for 1~ms, waiting for a L0 trigger decision; in the common system 
this is done on the TEL62 boards (section \ref{sec:tel62}) and their daughter-cards (section \ref{sec:daughters}). 
During such latency, the L0 trigger is elaborated on the full information from participating sub-detectors, producing data (L0 trigger 
primitives) indicating the fulfilment of several programmable criteria at the same rate (sections \ref{sec:tdcl0}, \ref{sec:lkrl0}) at which 
events occur. 
Such processing is performed in a time-asynchronous way, thus allowing to exploit the same cheap packet-based network protocol used for data 
readout.

The sub-detectors participating to the L0 trigger decision are currently: two plastic scintillator hodoscopes (NA48-CHOD, with a bar layout, 
and CHOD with a pad layout) and a Ring-Imaging \v{C}erenkov detector (RICH), both mainly used for timing and track-identification; 
coarse-grained digital data\footnote{Coarse-graining is required because of the sheer amount of electro-magnetic calorimeter data, from 13 
thousand 40~MHz continuously digitized channels, which cannot be made fully available within the time constraints of L0, see section 
\ref{sec:lkrl0}.} from cells of the electro-magnetic (LKr) and hadronic (MUV1, MUV2) calorimeters; lead-glass ring-shaped detectors surrounding 
the decay region (LAV) and small auxiliary detectors (SAC, IRC) for vetoing photons; and a plane of plastic scintillator pads behind an iron 
wall (MUV3) for vetoing muons. Data from the above detectors provide the required factor 10 rate reduction.
L0 trigger primitives are partially time-ordered, grouped into Multi-Trigger Packets (MTPs) and sent through one GbE link per sub-detector to 
the central L0 Trigger Processor (L0TP). These primitives are also read-out independently for monitoring purposes (section 
\ref{sec:primitive_readout}).

The L0TP (section \ref{sec:l0tp}) time-matches and processes the above primitives to concurrently check several programmable L0 trigger 
conditions.
Inclusion of more sub-detectors is possible, to further refine the trigger conditions or to collect alternative data samples.
L0 trigger primitive generation is required to occur within a programmable, fixed time limit (up to several hundreds of $\mu$s) after the event 
time, in order to allow the L0TP to make its decision based on all the available information. To ease the time matching task of the L0TP, MTPs 
are expected to be delivered roughly (because of variable network latencies) every 6.4~$\mu$s.

The L0TP issues a L0 trigger decision which eventually drives the data transfer to the PC farm. Actually, the two most data-heavy sub-detectors 
only save event data onto secondary (longer-latency) internal buffers, and send it to the PC farm only in case of a subsequent positive L1 
trigger signal.

The L0 trigger decision is issued in a synchronous way\footnote{In this context ``synchronous'' denotes a signal occurring in a precisely 
defined 25~ns time-slot with respect to its originating cause, in this case the physics event in the detector.}, thus allowing a simplification 
of the trigger distribution network, which shares links with the clock and timing distribution network. 
Each L0 trigger carries a timestamp with 25~ns granularity, identifying which data should be transferred to the PC farm, in time windows  
whose size can depend on the individual sub-detector; a trigger-type qualifier is also dispatched, which allows both a different data 
handling for different trigger classes (including \emph{e.g.} calibration and monitoring triggers), as well as the use of the very same L0 
trigger path for broadcasting commands related to the functioning, synchronization, integrity and flow control of the entire TDAQ framework, 
with a further unification of links.
The timestamp associated to each L0-triggered event is defined by the L0TP, and for each burst it is in unique relationship with a 
sequential event number. The consistency between timestamp and event number is checked for all data (the timestamp being part of the event 
structure at all levels of data transport), as any mismatch would indicate an unacceptable trigger loss in part of the system, thus leading 
to data rejection.

Since vetoing efficiency and avoidance of undetected readout failures is crucial for the experiment, all sub-detectors must provide a response 
to each L0 trigger, even if there is no data to be transferred. Furthermore, such responses always contain identifiable data structures 
originating in each individual readout board. The above implementation intrinsically provides the required control on the ``live state'' of all 
sub-detectors in each event.
Another potentially dangerous misbehaviour of the TDAQ system would be a time mis-alignment between data from different sub-detectors,  
resulting in data containing information belonging to different triggers. This is avoided by periodic time-alignment checks and event-by-event 
timestamp-matching checks: all digital systems run on the same synchronous 40~MHz clock, thus allowing locally- and centrally-generated 
timestamps to be compared. Furthermore, each individual electronic board records the number of 25~ns clock periods counted during each burst, 
and such numbers are compared in order to identify any loss of time synchronization. 

The unified data-path approach was pursued by extending it to the collection of ``slow-control'' monitoring information from the entire TDAQ 
system: at times (most notably at the end of each burst) special L0 triggers are dispatched, to which all boards react by sending monitoring 
data ``events'' along the standard data links. This approach does not require additional slow-control data paths, and ensures the availability 
of monitoring data together with the main data without needing additional book-keeping or data handling.

An extension of the system based on the hard real-time use of GPUs is described in section \ref{sec:gpu}.
Section \ref{sec:running} illustrates some results from the experience gained in running the system during the first data-taking period of 
the experiment, and some conclusions are presented in section \ref{sec:conclusions}.

\section{Common infrastructure}
\label{sec:common}

\subsection{Clock and L0 trigger distribution}
\label{sec:clock}

A common master clock signal with a $\sim$40~MHz frequency is generated by a single free-running high-stability oscillator\footnote{Hewlett 
Packard 8656B Signal Generator.} and optically distributed to all systems through modules of the Timing, Trigger and Control (TTC) system, 
designed at CERN for LHC experiments \cite{TTC}.
The master clock drives the entire TDAQ system and is used as the reference for all time measurements in NA62. While the common experiment 
timestamp is defined by such phase-coherent distributed clock, each sub-system locally generates by multiplication a properly locked reference 
for fine-time.

The master clock frequency is actually 40.078~MHz, since it must fall within the locking range of the QPLL (Quartz crystal Phase-Locked Loop) 
jitter-cleaning system \cite{QPLL} of the TTC system, used to guarantee the required clock accuracy and stability (such range was 
defined by the timing structure of the LHC machine). As a consequence all references to \emph{e.g.} ``25~ns'' should be understood to be 
the period of the main clock, close to 24.951~ns, and similarly the ``100~ps'' fine-time unit is actually 97.466~ps\footnote{The exact 
frequency of the master clock is irrelevant for NA62, as long as it is constant throughout the system.}.

The master clock signal is distributed to a fan-out card, which drives in parallel 12 identical clock/trigger sub-systems.
Each clock/trigger sub-system, normally serving a single sub-detector, comprises a modified version of the Local Trigger Unit (LTU) module 
\cite{LTU} designed for the ALICE experiment, and a TTC laser encoder and transmitter module (TTCex) \cite{TTCex} with up to 10 identical 
optical outputs. 
Passive optical splitters provide up to 320 output links per sub-system, to individually feed all boards.
Each electronics board requiring reference to the common experiment time is equipped with a TTC receiver (TTCrx) \cite{TTCrx} chip decoding 
information from the optical signal, and optionally a QPLL system to reduce clock jitter. 

All clock counters are simultaneously reset at the start of each burst, using a synchronous Start Of Burst signal (SOB) sent to all 
sub-systems through the TTC link before the arrival of the beam\footnote{This is generated by time aligning ``to 25~ns precision'' the SPS 
Warning of Warning of Extraction (WWE) signal, which is issued roughly 1 second before the first beam particles are delivered.}. 
This signal also defines the origin of the time measurements for the current burst. An analogous synchronous End Of Burst (EOB) signal is sent 
in the same way about 1~s after the end of the burst, defining the largest possible timestamp, whose value is recorded by each system and sent 
to the PC farm for logging together with the data. This allows (on-line and off-line) consistency checks on the number of clock cycles counted 
by each system during each burst. 
Note that by resetting all local timestamp counters on SOB through the same link which delivers the clock, any relative delay between 
sub-systems due to propagation time differences is irrelevant. 
Each sub-detector readout system is capable of running in a standalone mode, autonomously generating its own TTC signals (possibly including L0 
triggers) for test purposes, while during data-taking it runs under global experiment control.

When a L0 trigger is generated by the L0TP, a single optical link per board is used, via TTC time-multiplexing, to broadcast it to the rest of 
the experiment. The broadcast happens via two consecutive signals: the first is a ``L0 trigger'' pulse, the time of which defines the L0 
trigger time to 25~ns precision; the second encodes a 6-bit ``L0 trigger type'', which is asynchronous with respect to the L0 trigger pulse.
The TTC system imposes a 75~ns minimum time separation between two different L0 triggers, which is not an issue in principle, as the time 
occupancy of sub-detectors, and thus the chosen time widths of the readout windows, are larger than such figure.

The LTU provides the interface between the L0TP and the sub-detectors.  
The LTU receives signals from the L0TP, encodes and serializes them, and sends the data to the sub-detectors through the TTC system. 
It can also run in a stand-alone mode, in which it can emulate the generation of L0TP triggers, allowing each sub-detector to work independently 
during a debugging or calibration phase. 
The LTU also processes CHOKE/ERROR flow-control signals (section \ref{sec:flow}) from sub-detectors, and propagates them to the L0TP, where 
they are processed. 
The LTU also distributes SOB and EOB signals to all sub-detectors via the TTC system\footnote{These signals are encoded into the two lowest 
bits of the asynchronous TTC message byte, also containing the L0 trigger type. Such bits have defined reset behaviour for the TTC signal 
receivers, and are received at the same time by all boards within the TDAQ system.}.

The LTU provides the possibility to measure the phase of an incoming signal via an integrating RC circuit and an analog-to-digital converter, meaning appropriate delays for input signals can be applied to ensure proper time-alignment and latching.

\subsection{Flow and error control}
\label{sec:flow}

Two system-wide communication lines, named CHOKE and ERROR, are used by the TDAQ boards to provide feedback to the L0 trigger system concerning 
the occurrence of anomalous conditions which can impact the data-taking. Each board actively drives one CHOKE line and one ERROR line.
All the CHOKE (ERROR) lines from the boards of a given sub-detector are OR-ed together by dedicated active fan-in boards (CHEF), until there 
is a single pair of lines from the sub-detector. The pair of lines is then connected to the L0TP (section \ref{sec:l0tp}), via the sub-detector 
LTU, using point-to-point LVDS signals.

The CHOKE signal is used to indicate that (part of) a sub-system is approaching a state in which it will no longer be able to correctly 
handle data, because \emph{e.g.} its processing or storage resources are almost saturated. If the asserted CHOKE from a sub-detector is not 
masked in the L0TP the dispatching of L0 triggers is stopped until all sub-detectors stop asserting the signal. The CHOKE signal is used to 
exert back-pressure from TDAQ boards in case of anomalous rate conditions, but its assertion is not associated to any critical malfunctioning 
or data loss. 
The ERROR signal is used instead to indicate that (part of) a sub-system actually lost some data because its processing capabilities were 
exceeded. 
Tight data integrity control is obtained by dispatching special L0 triggers whenever either the CHOKE or the ERROR condition is asserted 
or de-asserted (see section \ref{sec:l0tp}), and the mandatory replies to such triggers act as active acknowledgements by sub-systems, 
traceable in the data, that the above conditions were properly handled while such systems were fully working.

\subsection{Configuration}

The state of the whole TDAQ system is centrally managed by the Experiment Control System (ECS), which runs a finite state machine. 
Two stages are foreseen to configure the TDAQ system: at the \emph{initialization} stage a complete restart of the system is performed, 
uploading to the hardware all the configuration data which is not meant to change frequently, and which might require a relatively long set-up 
time (several minutes); at the \emph{start run} stage a faster warm start occurs, in which further run-specific configuration data is uploaded. 
ECS communication is handled through the DIM system \cite{DIM}.
All configuration data is contained in human-readable XML files which are extracted from a database by a Run Control system, and stored 
into a condition database for each run.
Most sub-systems also deliver some configuration data, together with monitoring information, within their End-Of-Burst data packet, making it 
readily available inside the event data files.

\section{Common TDAQ board}
\label{sec:tel62}

The guiding principles driving the design of the TDAQ system were large channel integration, scalability and versatility, in order to optimize 
the implementation and maintenance effort, allowing use in a large set of heterogeneous sub-detectors, while ensuring the possibility of 
significant changes in trigger configuration.
A high-performance, general-purpose, versatile trigger and data-acquisition board, denoted TEL62 \cite{TEL62}, was designed to be used by most 
sub-detectors in NA62, with enough flexibility to be suited for rather different daughter-cards (such as the TDCB time digitizer card, 
section \ref{sec:daughters}).
Most sub-detectors in NA62 use this common system for readout and (some of them) for L0 trigger primitive generation: the \v{C}erenkov kaon 
tagger (KTAG), the charged particle scintillating veto counter (CHANTI), the downstream plastic scintillator hodoscopes (CHOD and NA48-CHOD), 
the main \v{C}erenkov detector (RICH), the muon (MUV0, MUV3) and hadron (HASC) veto counters, the photon veto counters (LAV, SAC, 
IRC\footnote{The latter two being also read-out by the calorimeter system.}).
Some sub-detectors adopted instead other dedicated systems, either because of high-density sensor integration requirements for the silicon pixel 
GigaTracker beam spectrometer, or the preference for a cheaper FPGA-based time digitizer solution, best suited to the main spectrometer's straw 
chambers reduced intrinsic time resolution, or the use of continuously digitizing flash ADCs for the electro-magnetic (LKr) and hadronic 
(MUV1, MUV2) calorimeters, which still use TEL62 boards for L0 trigger primitive generation.

\subsection{TEL62 hardware}
\label{sec:tel62hw}

\begin{figure}[hbt]
  \centering
  \includegraphics[width=\linewidth]{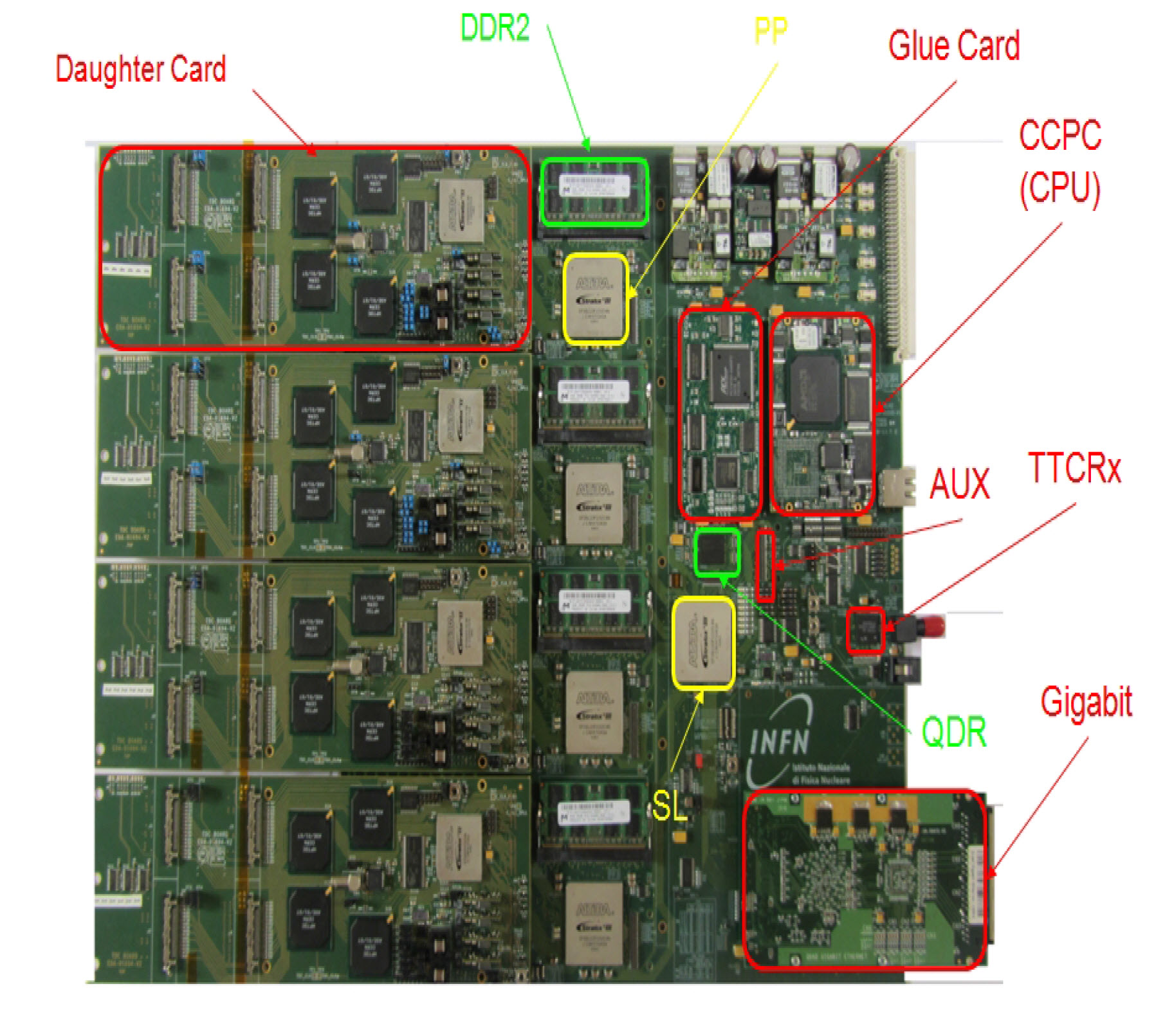}
  \caption{TEL62 board equipped with four TDCBs.}
  \label{fig:tel62}
\end{figure}

\begin{figure}[hbt]
  \centering
  \includegraphics[width=0.9\linewidth]{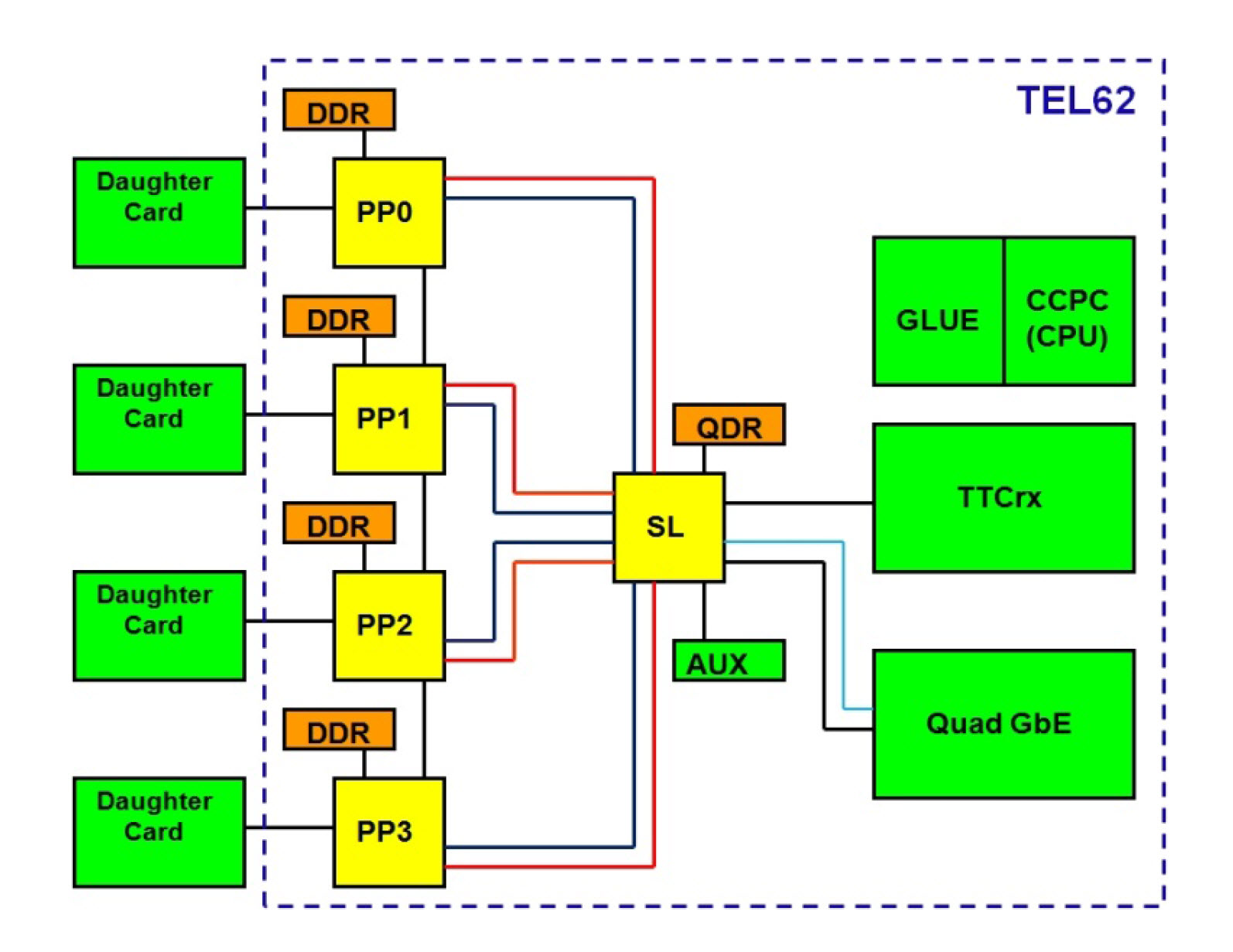}
  \caption{TEL62 block diagram.}
  \label{fig:tel62arch}
\end{figure}

The TEL62 board (fig.~\ref{fig:tel62}) has a similar overall architecture to the TELL1 board developed for the LHCb experiment\cite{TELL1}, 
but is based on much more powerful and modern devices, resulting in more than four times the processing power and more than twenty times the 
buffer memory of the original, with several other improvements in terms of connectivity.
Overall, about 100 TEL62 boards were produced, most of them being actually installed in the experiment.
The architecture of the TEL62 is shown in fig.~\ref{fig:tel62arch}.

Each of 4 FPGAs\footnote{Altera Stratix\reg~III EP3SL200F1152C4 with 200,000 logic elements and 9~MB embedded memory.} (Pre-Processing or 
``PP-FPGA''), is connected to a daughter card through a high density Samtec 200-pin connector, and to a 2~GB DDR2 memory buffer (in SO-DIMM form 
factor) which stores data during the L0 trigger latency. Each PP-FPGA is also connected to its neighbour(s) by two uni-directional 16-bit 
buses, which can be used for daisy-chaining.

A central FPGA of the same type (Sync-Link or ``SL-FPGA''), is connected to each PP-FPGA by two independent 32-bit data buses, for 
data and L0 trigger primitive flows respectively; each bus runs at 160~MHz and all signal lines are equal in length. 
The data and L0 trigger primitives from all PP-FPGAs are merged on the SL-FPGA, then formatted and stored in buffers: that for data is based on 
a Quad Data Rate (QDR) synchronous dual-ported SRAM, whose high bandwidth allows simultaneous read and write operations. With a 16-bit bus at 
double data rate and 100~MHz clock frequency, a bandwidth of 3.2~Gbit/s is reached. The chosen QDR device\footnote{Samsung Semiconductor 
K7Q161852A-16.} has a depth of 1~MB. 

The SL-FPGA is also connected by another 120~MHz bus to an output daughter card, through which data is eventually sent to other parts of the 
TDAQ system. The default output card is a custom quad-Gigabit Ethernet card (Quad-GbE) \cite{QuadGbE} developed for the LHCb experiment, which 
implements\footnote{Using the Intel IXF1104 as Media Access Control and Marvell Alaska 88E1140 as physical controller devices.} 4 $\times$ 
1 Gbit copper Ethernet channels.

The slow control, monitoring and configuration of the TEL62 is handled by 2 more daughter cards inherited from the TELL1 design: a 
commercial Credit-Card PC (CCPC) running Linux\footnote{CERN Scientific Linux 4 \cite{ScientificLinux}.} and a custom I/O interface card 
(``Glue Card'') connected to the CCPC through a PCI bus. 
Three different communication protocols are implemented on the Glue Card and distributed to all devices and connectors on the TEL62: JTAG, 
I2C and ECS. JTAG is used to remotely configure all the board devices, I2C is mainly used to access registers on some daughter cards, 
while ECS is a custom protocol used to access the internal registers of the PP- and SL-FPGAs via a 40~MHz 32-bit bus.

The 40~MHz experiment clock and the L0 trigger information are distributed to TEL62 boards through a CERN-standard optical TTC link. 
The TEL62 uses a TTCrx chip to decode the clock and trigger information. The clock signal is sent to the SL-FPGA and distributed from there to 
the PP-FPGAs through internal PLLs that upscale it by 4 and set the correct phase to correctly latch incoming data. 
A further layer of communication is provided by an auxiliary connector, which can be used to plug in an inter-communication card for 
daisy-chaining several TEL62 boards (section \ref{sec:intertel}).

The board size complies with the 9U Eurocard standard (340 x 400 mm). It gets power from a VME-like connector (not used for communication) 
compatible with that of the TELL1. Overall power consumption of the TEL62 board is around 50 W.
The printed circuit is made of 16 layers, with all lines controlled in impedance (50 $\Omega$). Special care was taken in routing the clock 
tree, to minimize time jitter, and in equalizing the lengths of data bus lines. 

\subsection{Common TEL62 firmware}
\label{sec:generic_fw}

Appropriately for a multi-purpose TDAQ board, a large part of the firmware design is common to all TDC-based sub-systems and is described 
in this section, while sub-detector specific parts are detailed in subsequent sections.

The common part of the firmware consists of about 65,000 user-written lines of VHDL code.
The firmware is hierarchically managed with common and sub-detector specific libraries through the Mentor Graphics HDL 
Designer\reg~software suite, interfacing to the Altera Quartus II\reg~compiler, and to the Apache Software Foundation SVN\reg~versioning 
system, with a central repository allowing concurrent development by the various institutions participating in the project.

Multiple clocks are used within the firmware, with most of the modules running on a 160~MHz main clock, locally generated inside each FPGA 
and phase-locked to either the common experiment-wide 40~MHz master clock received from the TTC (the SL-FPGA) or to a 40~MHz clock 
distributed by the SL-FPGA (each PP-FPGA, with programmable individual phase-adjustment).
Clock frequency constraints for some external devices make the use of different clock domains within FPGAs unavoidable, and care had to be 
taken in order to control timing-closure violations.

The UDP protocol was chosen as a light-weight solution for output data transmission over the GbE links, allowing direct connection to a standard 
switched network. This choice was dictated by requirements of simplicity and high throughput. To deal with the unreliable UDP protocol, which 
has no re-transmission, extensive error detection was implemented.

Quite extensive test and debugging features are implemented in the firmware, as required to control a rather complex system. These are 
similar within all FPGAs and include, besides a large number of user accessible registers and the capability to read and write internal buffer 
memories and FIFOs by the CCPC, the presence of  a ``freeze'' logic to halt all processing, which can be triggered by the occurrence of 
programmable conditions (errors, timestamp counts, buffer filling, etc.). A distributed ``logging'' system is implemented as a shared 
internal memory to which most firmware modules can write a variable number of timestamped data words to report the occurrence of specific 
conditions. Such memories can be read by the CCPC, and their use, by selectively masking through registers the firmware modules and message 
severity levels which to be logged, was very useful during commissioning, to monitor and understand rare/anomalous conditions.

Configuration, slow control and monitoring of the board (and its daughter-cards) is performed by the CCPC. 
A versatile management program with command-line interface and extensive scripting and macro capabilities was developed, and is used 
for all communication with the board; it consists of about 55,000 lines of C code, interfacing to low-level hardware libraries. 
An interactive version is used for testing and monitoring, while during data-taking the program runs as a daemon and communicates through DIM
with the TDAQ control system for board configuration, error checking and monitoring.

\subsubsection{PP-FPGA logic}
\begin{figure*}[hbt]
  \centering
  \includegraphics[width=0.9\linewidth]{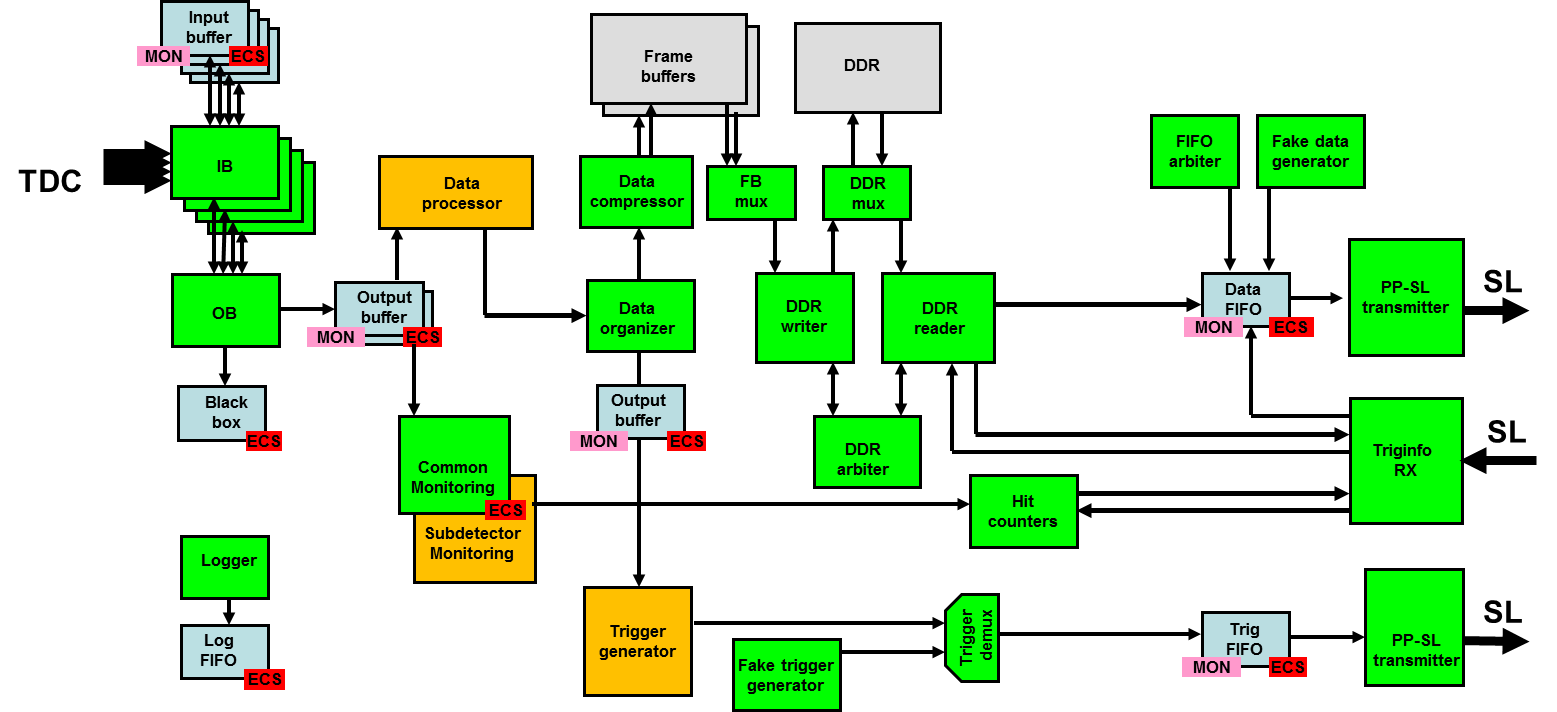}
  \caption{Schematics of the common firmware in the TEL62 PP-FPGAs. Buffers whose filling is continuously monitored to act on flow control are 
  labelled ``MON''. Buffers which are accessible by the CCPC for monitoring and debugging purposes are labelled ``ECS''.}
  \label{fig:Generic-PP}
\end{figure*}

A schematic of the common PP-FPGA firmware (the same for all four devices) is shown in fig.~\ref{fig:Generic-PP}.
The common PP-FPGA logic is configured by more than 100 32-bit registers, and most of it runs on a 160~MHz clock. 

Each PP-FPGA handles data received from a daughter-card, which in the case of TDCBs arrives as independent 32-bit parallel data streams from 
each of four TDC chips. The data from each TDC chip is stored into dedicated 2K word deep Input Buffer (IB) FIFOs, and monitored on-the-fly in 
order to identify corrupted data, parity errors, malformed data frames or repeated words. These issues indicate upstream errors, and cause 
appropriate error flags to be set and later transmitted together with the data.
The IBs act as de-randomizing storage and provide the input to the following merging stage; the latter patches together the individual TDC data 
words belonging to the same 6.4~$\mu$s-long data frame (identified by a leading timestamp word) and produces a merged data frame with updated 
word count field, to be stored into a 2K word deep Output Buffer (OB) FIFO. A copy of the merged data is written into another identical 
buffer feeding data monitoring modules.
Furthermore, half of the OB data is constantly available in a circular buffer accessible from ECS, useful for \emph{post-mortem} 
debugging in case a fatal error condition triggers the freeze logic.

After an optional sub-detector specific data processing module, meant for data calibration, remapping or processing to be performed on-the-fly, 
data frames are passed to a Data Organizer (DO) module. 
The DO unpacks data from each time frame and arranges them into 256 time-ordered slots, each time slot being 25~ns long.

The large L0 trigger latency and high data rate (which is sensitive to beam rate fluctuations) require a sizeable amount of internal RAM devoted 
to storing data in the 256 time slots belonging to each data frame. Moreover, in order to cope with the continuous data flow without introducing 
any dead time, two identical memories (32K words each) are used, so that one time frame is unpacked while the previous one is being 
processed by the following firmware modules.

The Data Compressor (DC) module compacts the data for one 6.4~$\mu$s time frame, allowing a variable number of words in each 25~ns time 
slot, in order to optimize its storage into the external memory, thus minimizing the required number of accesses when data is written 
and read back from it.

Such external DDR memory is organized into 64 million 256-bit wide locations per PP-FPGA. Half of the memory space is used as a circular buffer 
to store data for 131,000 data frames, a latency in excess of 800~ms before overwriting, each frame allowing up to 2K 32-bit words, 
corresponding to a 80~MHz rate of data words from a single TDC chip (which can steadily produce slightly less than 40~MHz of data words at most). 
The other half of the DDR memory is used to store the information about the dynamically defined starting address and number of data words for 
each 25~ns time slot: addresses for up to 1 billion time slots are stored, corresponding to 26.8 s worth of data.

Access to the DDR memory occurs in pages of up to 32 256-bit words, through an ALTERA proprietary firmware driver running at 320~MHz, which takes 
care of memory refresh operations, while arbitration between the periodic data frame writing process (6.4~$\mu$s period) and the aperiodic 
reading process (at the L0 trigger rate, up to a maximum of once per $\mu$s) is taken care of by the firmware.
The CCPC can also access indirectly the DDR memory in blocks of 1 kB, for test or debugging purposes.

The sub-detector-specific Primitive Generator modules (for sub-detectors participating to L0 trigger generation) also receive a copy of the 
frame-merged and time-ordered data from the DC, through a dedicated 2K word deep Trigger Input Buffer (TIB), and they process it on-the-fly to 
produce L0 trigger primitives; such modules are described in section \ref{sec:tdcl0}.

Upon reception of a readout request from the SL-FPGA (corresponding to a L0 trigger) each PP-FPGA sends the formatted data corresponding to a 
programmable time window around the trigger time to the SL-FPGA.
The PP-FPGA reads the relevant trigger information (25~ns trigger timestamp and 6-bit trigger type) from a dedicated trigger information FIFO, 
and the Trigger Information Receiver (TRIGINFORX) module handles it accordingly. For triggers involving real data (such as ``physics'' triggers) 
the DDR Reader module receives the trigger time stamp and the (programmable, possibly depending on trigger type) number and offset (with 
respect to the trigger time) of the time slots to be read; it reads from the DDR the corresponding data, which is formatted and written in the 
2K word deep final data FIFO buffer. 
For special triggers (for which actual TDC data is not required) the TRIGINFORX fills the final data FIFO itself. In the case of End Of Burst 
triggers, write access to the final data FIFO is granted to other firmware modules, such as a hit-counting module and other sub-detector 
specific modules, which sequentially append their monitoring data.
The data is then moved from the final data FIFO to the SL-FPGA, with checks on parity and event size.

Both the data and L0 trigger primitive paths from PP-FPGA to SL-FPGA can be tested using data words produced by embedded pseudo-random bit 
sequence generators; this allows the best relative phase delay between the two FPGA clocks for error-free communication to be determined, as 
such delay can vary depending on the actual version of the firmware loaded into the FPGA, due to the internal routing of the logic by the 
compiler.
Overall, the common part of the firmware uses 55\% of the PP-FPGA logic resources and 45\% of the internal memory.

\subsubsection{SL-FPGA logic - data}
The SL-FPGA logic handles two separate data flows: one for the main event data being read after a L0 trigger and sent to the PC farm, and one 
for the L0 trigger primitives being continuously produced and sent to the L0TP, the latter being present only for sub-detectors participating 
to L0 trigger generation.
 
\begin{figure*}[hbt!]
  \centering
  \includegraphics[width=0.9\linewidth]{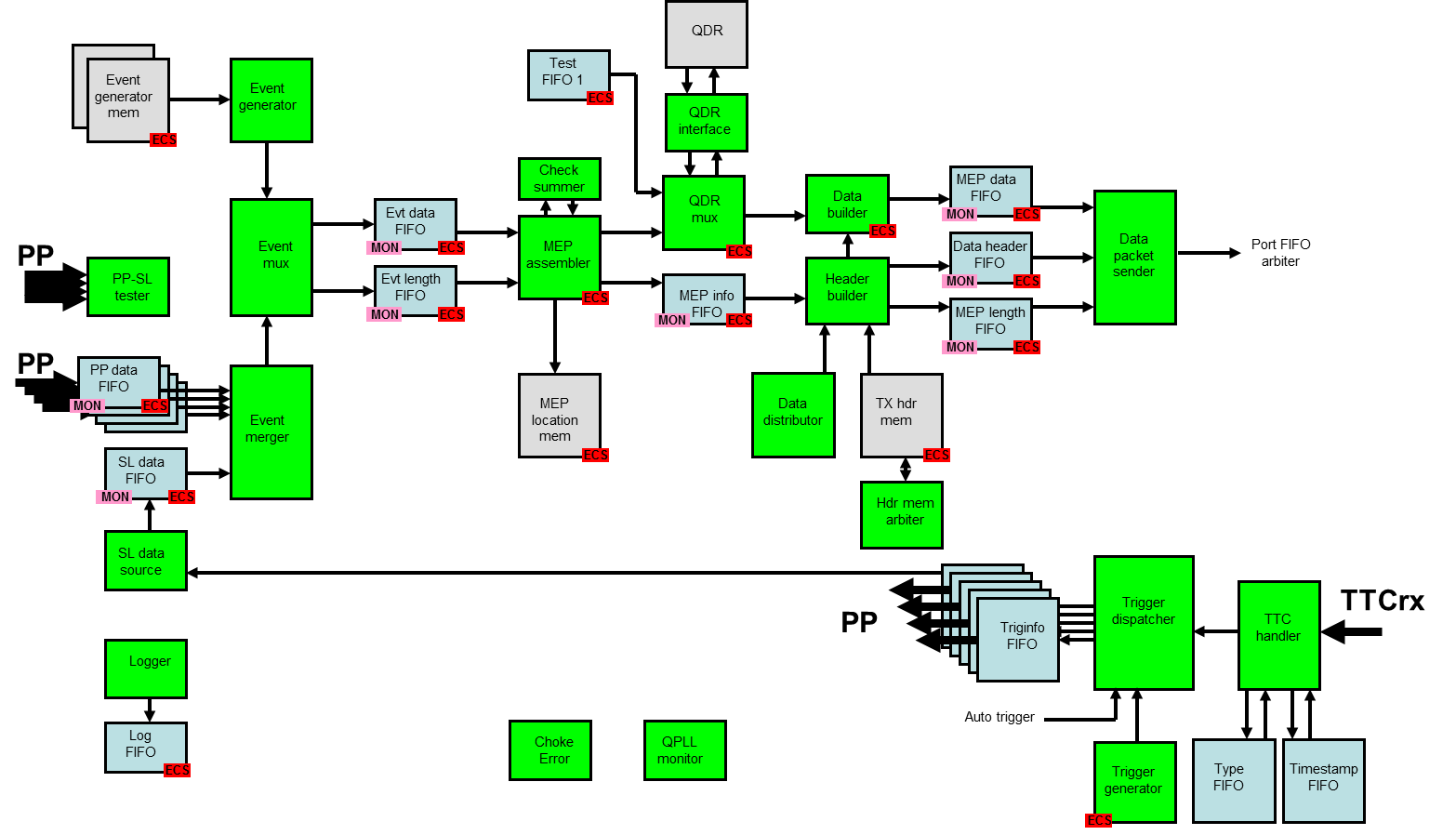}
  \caption{Schematics of the common firmware for the data path in the TEL62 SL-FPGAs. Buffers whose filling is continuously monitored to act on 
  flow control are labelled ``MON''. Buffers which are accessible by the CCPC for monitoring and debugging purposes are labelled ``ECS''.}
  \label{fig:Generic-SL-data}
\end{figure*}

The main data flow is described first and is shown schematically in fig.~\ref{fig:Generic-SL-data}.
The non-subdetector-specific SL-FPGA logic is configured by about 150 internal 32-bit registers; the logic mostly runs on a 160~MHz clock, 
except for the interface to the output GbE links which is limited to 120~MHz by the external devices, thus requiring separate clock domains.

The first part of the SL-FPGA logic merges event data, which was received from the four PP-FPGAs and stored in 2K word deep FIFO buffers, 
together with locally produced data, and stores the complete event into a buffer FIFO. For testing purposes fake events can be read from an 
embedded RAM instead.

The MEP Assembler module arranges events into Multi Event Packets (MEPs), containing a programmable number of events, in order to optimize 
network transmission bandwidth. The module formats the data, while also computing and appending a CRC checksum to verify integrity at later 
stages. MEPs are then stored into the external QDR memory arranged as a circular buffer, where they lie waiting for further encapsulation and 
transmission. The location of the most recent 256 MEPs is recorded in a dedicated memory for \emph{post-mortem} debugging. 

The following Packet Builder stage extracts MEPs from the QDR memory and encapsulates them into network packets, by adding the UDP, IP 
and Ethernet protocol headers. Jumbo Ethernet frames are supported, as well as IP fragmentation.
All sub-detectors must send data fragments corresponding to the same event to the same PC of the PC Farm: this requires event distribution to 
be fully coherent among all sub-detectors, and has implications on the number of events stored in a MEP, the number of output GbE ports used 
and the packet addressing to farm nodes. 
A flexible three-level round-robin distribution mechanism for MEPs is implemented. Each of the four TEL62 GbE ports has a list of up to 63 
programmable destination IP addresses over which it cycles, and a programmable number of MEPs is sent to one port before switching to the 
next one. 
Independently from the above cycle, the destination address for the current port is changed to the next one in the list after a (distinct) 
programmable number of MEPs is sent.

\begin{figure}[hbt!]
  \centering
  \includegraphics[width=\linewidth]{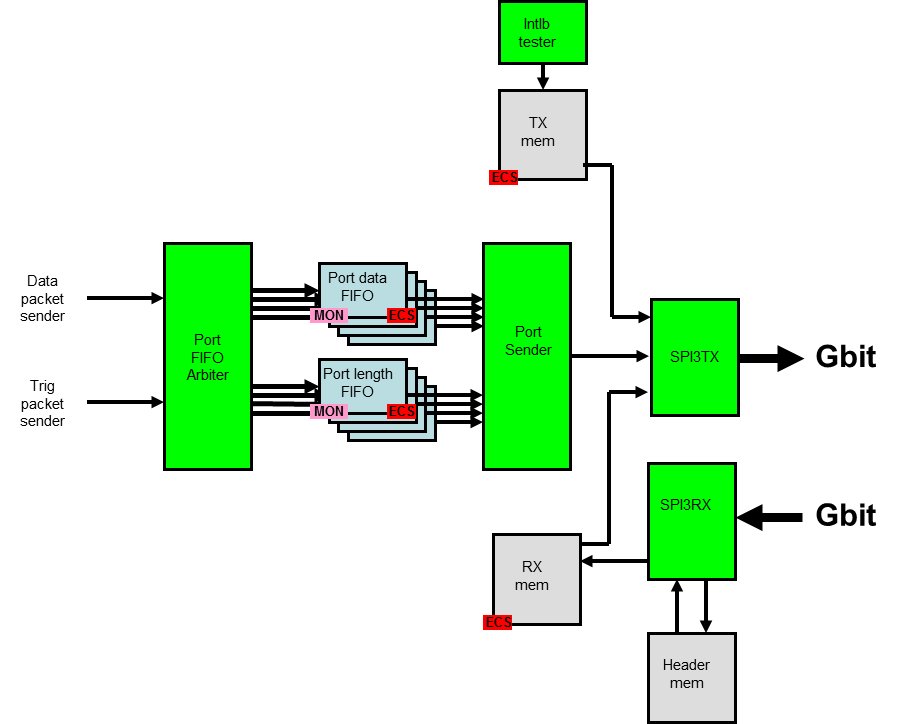}
  \caption{Schematics of the common firmware for the output section of the TEL62 SL-FPGAs. Buffers whose filling is continuously monitored to 
  act on flow control are labelled ``MON''. Buffers which are accessible by the CCPC for monitoring and debugging purposes are labelled ``ECS''.}
  \label{fig:Generic-SL-out}
\end{figure}

The following part of the logic, shown in fig.~\ref{fig:Generic-SL-out}, is shared by both the main event data and the L0 trigger primitive 
flows. Each of the four output GbE ports can be configured to be dedicated to either of the two flows.

Formatted packets destined to a given output port are stored into 8K word deep buffer FIFOs (one per port), allowing to decouple the flow 
from any temporary congestion of individual hardware links.
The Port Sender module actually transfers the packets to the Quad-GbE card (whose input interface is shared by all ports) in an order which 
depends on the current availability of a port.
Dedicated modules allow using the output links to either transmit pre-programmed data from an internal memory, or to store the data received 
from the output links when configured in mirroring mode, and are used for board tests. 

The logic to handle L0 triggers comprises a TTC Handler module which timestamps the triggers received from the TTC network and stores them 
into a FIFO buffer, similarly storing trigger type words and pairing them with the corresponding trigger timestamp values according to their 
order of arrival. Triggers are then delivered to each PP-FPGA by the Trigger Dispatcher module.
Debugging and test features include a circular buffer storing the last 256 triggers dispatched, the possibility to autonomously generate 
triggers when specific primitive data word patterns are recognized, and pre-programmed trigger sequences under CCPC control.

\subsubsection{SL-FPGA logic - L0 primitives}
\begin{figure*}[hbt]
  \centering
  \includegraphics[width=0.9\linewidth]{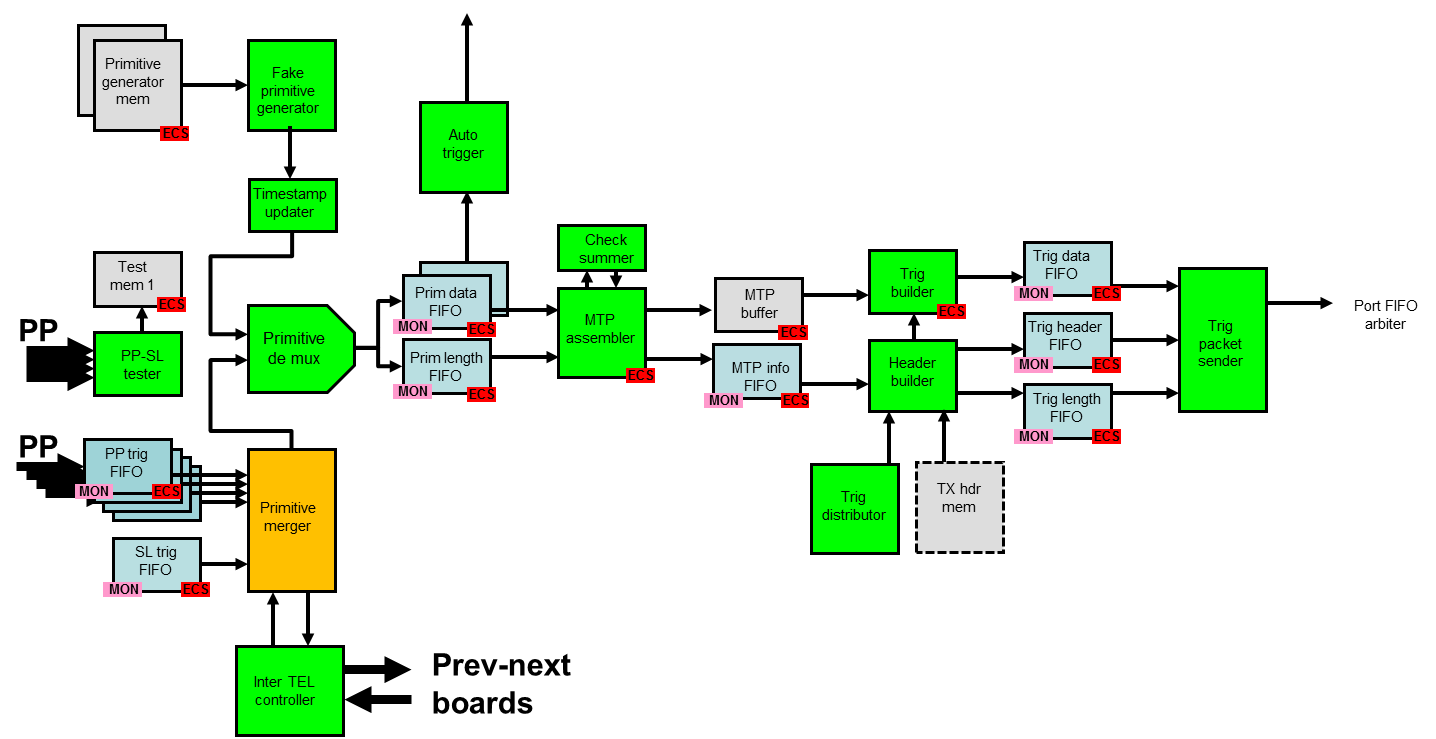}
  \caption{Schematics of the common firmware for the L0 trigger primitive path in the TEL62 PP-FPGAs. Buffers whose filling is continuously 
  monitored to act on flow control are labelled ``MON''. Buffers which are accessible by the CCPC for monitoring and debugging purposes are 
  labelled ``ECS''.}
  \label{fig:Generic-SL-trig}
\end{figure*}

For sub-detectors involved in L0 trigger generation, the flow of L0 trigger primitives proceeds in parallel to the main data flow, and is shown 
schematically in fig.~\ref{fig:Generic-SL-trig}.
Overall, the common part of the firmware uses 20\% of the SL-FPGA logic resources and 35\% of the memory.

Primitive fragments from enabled PP-FPGAs are merged by a sub-detector-specific module. Fake primitives can be generated within the SL-FPGA 
for test purposes, being read from a dedicated memory and with their sequential number and timestamp being changed on-the-fly to resemble 
real primitives.
Merged primitives are then aggregated into Multi Trigger Packets (MTPs) by the MTP Assembler, and temporarily stored in an internal 4K-word 
deep circular buffer: the smaller size of primitives, with respect to event data, does not require the use of a large external memory device. 
MTPs can either contain a fixed number of primitives (with a time-out mechanism to avoid excessive latency accumulation) or can be 
sent at periodic time intervals (6.4~$\mu$s period), which is the normal working mode required by the L0TP (see section \ref{sec:l0tp}).
As for the main data flow, the Trigger Builder and Sender modules handle the preparation of UDP packets and their transmission to the output 
links through the dedicated port buffers.
A test feature allows selectable trigger primitives to autonomously generate L0 triggers locally, to be used when the L0TP is not available.

\section{TEL62 daughter cards}
\label{sec:daughters}

Each TEL62 can host up to 4 daughter-cards, providing input data to be handled by the corresponding PP-FPGAs; while the TELL1 daughter-cards 
developed for LHCb are mechanically compatible with the TEL62, different cards were developed for NA62 to provide time measurements on 
digitized detector-signals (section \ref{sec:tdcb}), or to handle digital pulse-height information from calorimeters (section \ref{sec:lkrb}). 
Data output is provided by another daughter-card, which can be either the Quad-GbE card described in section \ref{sec:tel62hw}, or a transmitter 
card based on a custom protocol (section \ref{sec:lkrb}).
Furthermore, inter-connection cards were developed to allow TEL62s to share data among themselves (section \ref{sec:intertel}).

\subsection{TDC boards}
\label{sec:tdcb}

Most sub-detectors in NA62 exploit their good time resolution in order to cope with the high-rate of events: a time-digitizer card (TDCB) 
was developed to handle time information \cite{TDCB}.
Overall, more than 130 TDCB were produced, most of them being actually installed at the experiment.

The design of the TDCB was driven by the desire to integrate a large number of channels on a single card, in order to ease trigger generation. 
The quest for a compact and common electronics, and the relatively short distance between sub-detectors and a site where readout electronics 
could be placed (in absence of severe space constraints), led to the choice of having digitizers on the readout board itself. This left only 
analogue (and sub-detector specific) front-end electronics on each sub-detector in a potentially higher radiation environment, with most 
clocked digital components being close together, at the price of transmitting the pulses to be time-digitized over $O$(5~m) long LVDS cables.
The requirements of a good time resolution and high channel integration led to the choice of the CERN High-Performance Time-to-Digital 
Converter \cite{HPTDC} (HPTDC) as time digitizer.

Each HPTDC works in fully digital mode and hosts 32 TDC channels when operated in high-resolution mode (100~ps LSb), with some internal channel 
buffering for multi-hit capability and a trigger-matching logic allowing the extraction of hits in selected time windows. Channels are arranged 
in groups of 8 sharing some internal buffers.
While trigger-matching mode was implemented to allow the chip to act as front-end buffer, by storing digitized data while a trigger signal is 
generated, in NA62 the L0 trigger latency was chosen to be much longer than the maximum storage time allowed by the HPTDC before internal 
timestamp roll-over (51.2~$\mu$s), in order to allow complex trigger decisions to be performed at the lowest trigger level.
Moreover, in NA62 the L0 trigger is computed based on the acquired data itself.
For these reasons, the HPTDC is used in trigger-matching mode just as a way to obtain properly time-framed data from it, as explained in the 
following section.

Four HPTDC chips are mounted on each TDCB, for a total of 128 channels (512 channels per fully-equipped TEL62 carrier board); all TDC channels 
for most small sub-detectors are thus hosted on a single TEL62 board, and the entire 2000 channels of the RICH sub-detector are managed by just 
4 TEL62s.
The measurement of both the leading and trailing edge times allows analogue pulse-height information to be obtained by a time-over-threshold 
method, and HPTDCs can indeed be programmed to digitize the time of occurrence of both signal edges, provided they exceed a 7~ns time separation.

The TDCB houses four 68-pin VHDCI connectors for input signals, each of them delivering 32 LVDS signals to one HPTDC, with two spare pairs 
being used to provide additional grounding and to allow user-defined back-communication from the TDCB to the front-end electronics. This latter 
feature can be used to trigger the injection of calibration pulses in a sub-detector or calibration patterns in its front-end electronics. 
The connection between front-end and TDC chips can use standard SCSI-3 cables, or higher-performance ones if needed.

The TDCB houses a dedicated FPGA\footnote{Altera Cyclone\reg III EP3C120F780 with 120,000 logic elements.} named TDC Controller 
(TDCC-FPGA) which handles the configuration of the four HPTDCs, reads the data they provide, and optionally pre-processes it. 
A 2~MB external static RAM block is also present, and can be used for on-line data monitoring purposes and low-level data-quality checks.
The TDCC-FPGA can be configured by an on-board flash memory or through JTAG (either using a JTAG port on the TEL62, or its embedded 
CCPC processor). Communication between each TEL62 PP-FPGA (section \ref{sec:tel62}) and the corresponding TDCC-FPGA proceeds through a 
200-pin connector hosting four independent 32-bit single-ended LVTTL parallel data buses (one per HPTDC) and dedicated lines for synchronous 
commands and resets.
The TDCC-FPGA can also be accessed from the CCPC on the TEL62 board via a dedicated I2C connection for slow operations, such as access to 
internal configuration registers. The HPTDCs are configured via JTAG, with the TDCC-FPGA acting as the JTAG master: configuration data is sent 
to the TDCC-FPGA from the CCPC via I2C, and is then uploaded to the HPTDCs. Alternatively, both the TDCC-FPGA and the four HPTDCs can be inserted 
into a global JTAG chain, which also includes all TEL62 devices and can be driven by the CCPC.

The TDC contribution to the time resolution depends on the random jitter of the reference clock against which the measurement is performed. 
The master 40~MHz clock is distributed via the TEL62 to the TDCBs, where it can be configured to be processed by two more jitter-cleaning 
stages: an on-board QPLL and the internal PLL of the TDCC-FPGA. 
Detailed tests performed in different configurations showed that the level of the jitter, measured with Time Interval Error (TIE) at 50\% 
level, is below 20~ps.

\subsubsection{TDCB firmware}
The firmware for the TDCC-FPGA is common to all sub-detectors.
The HPTDC chips are configured to run in high-resolution mode (100~ps LSb), usually generating two 32-bit words per input signal, with 
19-bit leading-edge and trailing-edge time digitization.

HPTDCs are used in trigger-matching mode, storing measurements in internal buffers, from where those with times around a ``trigger'' signal can 
be later extracted. However, such working mode is only used as a way to obtain properly time-framed data, and extraction is periodically driven 
by the TDCC-FPGA with no relation to the L0 trigger.
The HPTDC time-matching extraction parameters are thus set to read-out all hits which occurred since the previous extraction. This results 
in reading all hit data (only once), while overcoming the limited time-digitization range of the HPTDC, with the TDCC-FPGA appending 
a timestamp to each data frame, unambiguously associating each hit to an ``absolute'' time since the beginning of the burst.
With the chosen frame period of 6.4~$\mu$s and a 400~ns LSb, the range of the 28-bit frame timestamp word exceeds the maximum duration of a burst.

The TDCC-FPGA independently processes data from each of the four HPTDCs through a dedicated 32-bit bus using a block-write protocol, reading one 
data word every 25~ns. It formats data into timestamped frames, also adding a trailing word-count, as well as optional error words in case 
anomalous data or conditions are detected. Two redundant chip identifier bits in each hit word are replaced with parity bits to allow off-line 
data integrity checks.
Besides handling HPTDC configuration, the TDCC-FPGA JTAG master controller is also used to read status information during running.

Several test and debugging features are present in the firmware.
Two different TDC data emulators are implemented: one generates programmable repeating patterns on selected channels, and another cyclically 
transmits pre-loaded data words from internal memory buffers. The latter is used during the TEL62 acceptance test, using specific patterns to 
stress the boards by emulating different detector rate conditions.
A fraction of the TDC data stream can be stored into the board static RAM during running, from where it can be accessed at the end of a burst 
for debugging and monitoring purposes.
The TDCC-FPGA can drive a spare output LVDS line of each HPTDC input cable to trigger the front-end boards for sub-detector calibration: 
front-end boards can inject signals into the TDC chips in response to such stimuli, thus allowing a test of the whole chain. Such calibration 
can be driven periodically by a programmable counter or upon command from the TEL62 carrier board.

\subsection{Calorimetric trigger boards}
\label{sec:lkrb}
The calorimetric L0 trigger works on digitized pulse-height data, from the digital sums of 4 $\times$ 4 cells (``super-cells'') of the 
electromagnetic liquid krypton calorimeter (LKr), and single-channel data from other calorimetric sub-detectors. 

The calorimetric trigger processor is a parallel system, composed of TEL62 boards configured as Front-End, Merger and Concentrator devices 
with different functions: \\
$\bullet$ Front-End boards receive digital sums from the calorimeter digitizing modules, and perform peak searches in space to determine the 
time, position and energy of each detected peak; \\
$\bullet$ Merger boards (only used for the LKr) receive trigger data from the Front-End boards and merge peaks into clusters; \\
$\bullet$ The single Concentrator board receives peaks or clusters from the calorimetric sub-detectors, counts them, computes separate sums 
for electro-magnetic and hadronic energy, and generates trigger primitives.

In order to handle the 864 LKr super-cell channels (plus the 20 channels from other sub-detectors), 37 TEL62 boards are used, for which several  
daughter-cards were developed: receiver cards (TELDES) are used for receiving input data, and paired transmitter (Cal-L0TX) and receiver 
(Cal-L0RX) cards are used to pass information between the TEL62s, in a tree structure. Additional daughter cards are used to allow the 
coarse-grained input data used by the calorimetric L0 trigger to be sent to the PC farm, where it can be used for HLT processing.

The calorimetric L0 trigger mostly uses a dedicated firmware for TEL62 FPGAs: more details will be made available elsewhere.

\subsection{Auxiliary cards}
\label{sec:aux_cards}

A few more daughter-cards were developed for the TEL62, used for specific purposes.

\subsubsection{Interconnection card}
\label{sec:intertel}
For sub-detectors with more than 512 channels using TDC boards, the information is necessarily split over more than a single TEL62 board. 
In order to allow triggering algorithms that require to correlate data from the whole sub-detector, a bi-directional communication card was 
developed.

\begin{figure}[htbp]
  \centering
  \includegraphics[width=0.9\linewidth, clip]{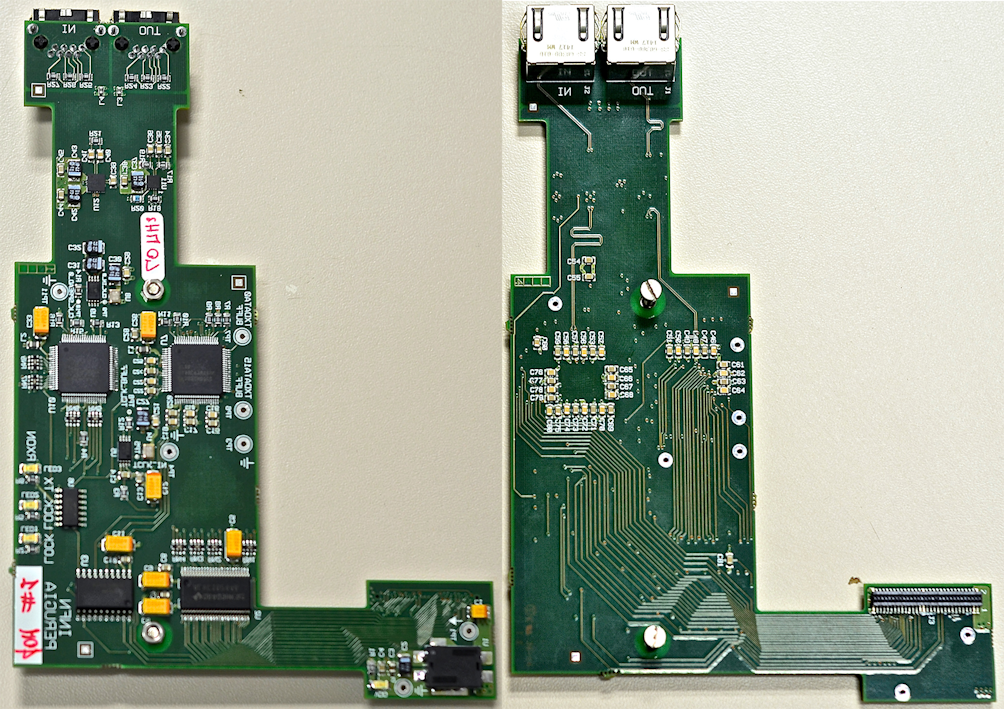}
  \caption{InterTEL board top (left) and bottom (right) views.}
  \label{fig:intertel}
\end{figure}

InterTEL cards (fig.~\ref{fig:intertel}) provide a daisy-chain link between different TEL62s \cite{intertelMB, intertelML}. 
They are connected to the carrier board by a 60-pin fine-pitch SMD connector, with buffered signals to reduce interference, noise, and 
cross-talk issues. 
The InterTELs are connected to each other via two RJ-45 connectors (one for TX and one for RX), with a LVDS bus for the physical layer and a 
proprietary serial communication protocol for the data link layer.

The protocol is implemented by serializer/de-serializer chips\footnote{Texas Instruments DS92LV16.} separately clocked at 40~MHz.
The data transmission rate of the link is 720~Mbps, including an overhead of 80~Mbps due to the embedded clock foreseen by the protocol.
The actual throughput is then 640~Mbps.
One twisted-pair of a Cat.~6 S/FTP 26-AWG cable is used, and a specific chipset\footnote{Texas Instruments DS15BA101 and 
DS15EA101.} is used to extend the maximum allowed cable length, so that the allowable 6~dB total link loss corresponds to 85~m.

\subsubsection{Pattern generation card}
A TEL62 daughter-card named PATTI, with level translators and VHDCI connectors was developed to implement a versatile digital pattern-generation 
system compatible with TDCBs, required for extensive system debugging and production testing. 
A dedicated PP-FPGA firmware controls the card, which can generate 128 LVDS signals in periodic patterns, or in a fully customizable way 
by reading individual channel data from an internal memory. The signal edges can be controlled to 6.25~ns precision.
In order to provide a fully self-contained test system, the card also provides a trigger output to emulate L0 triggers with no need for a 
dedicated LTU board (section \ref{sec:clock}), as well as CHOKE/ERROR signals for test purposes.
The PATTI board has the same format as the TDCB, and up to four can be housed on a TEL62, for a total of 512 output channels over 16 SCSI-3 
cables.

\subsubsection{Multi-purpose card}
Another TEL62 daughter-card named TALK was produced to interface parts of the old readout system of the NA48 experiment to the new TDAQ system 
during the first phase of NA62. The card works as a multi-purpose interface between a TEL62 and external devices: it houses the same FPGA used 
on the TDCB, one bipolar encoder/transceiver ``Taxi'' chip, 5 GbE links, 5 LEMO I/O connectors, 4 RJ-11 connectors and a multi-pin connector 
for the LTU card. The card is usually controlled from the TEL62 through dedicated parallel buses, but can also be accessed through dedicated 
JTAG and I2C connectors, and Ethernet as well.
The TALK board is twice the size of the TDCB, and up to two can be housed on a TEL62; a 6U VME frame was also developed for standalone use.
Besides its original use, it was successfully used as a L0TP emulator and calibration control system.

\section{TDC-data L0 trigger logic}
\label{sec:tdcl0}

As mentioned, L0 trigger primitives are sent in Multi Trigger Packets (MTPs), transmitted periodically for every 6.4us time frame, even if 
the MTP contains no primitives. Each primitive is coded into a single 32-bit word containing its ID (\emph{i.e.} the conditions which are 
satisfied), the 8 least significant bits of its 25~ns timestamp, and 8 fine-time bits (down to the 100~ps unit), thus identifying the primitive 
time up to 6.4~$\mu$s. The upper part of the timestamp, to cover the full duration of the burst, is stored in the header of each MTP. 
Up to 256 time-ordered primitives are stored in an MTP.
For the expected event rate at full beam intensity, the average maximum of 80 primitives per frame results in packets well below standard 
Ethernet payload limits. The corresponding average bandwidth is significantly less than 1 Gb/s even for the most active sub-detector: even 
allowing for rate fluctuations, at most one GbE link is used for MTPs, leaving three available for main readout data.

The L0 trigger primitive generation firmware is implemented in the TEL62 FPGAs, which also contain the common logic to handle the data flow and 
L0 trigger response (section \ref{sec:generic_fw}). Sub-detector channel data are correlated first in the PP-FPGAs (dealing with one TDCB 
each) and later in the SL-FPGA (dealing with the entire board).
A third layer of correlation logic might be implemented in the SL-FPGA in case more TEL62s are connected in a daisy chain, for sub-detectors 
with more than 512 TDC channels involved in the L0 trigger (section \ref{sec:intertel}).

\subsection{NA48-CHOD and RICH L0 trigger}
\label{sec:RICH}
The purpose of this firmware\cite{RICH_ref} is to produce clusters of hits belonging to the same event based on the hit times, providing a 
precise time reference and a hit count. 
No detector-specific information is used so that the firmware can work for different sub-detectors, such as RICH and NA48-CHOD. 
The main guidelines for the design were the minimization of resource usage and a stable latency for production of the primitives. 
The capability to cope with the full expected hit rate while maintaining as much versatility as possible were also followed as guidelines.
The firmware was designed to be reliable, adaptable to any primitive-generating sub-detector, easily upgradable, and compatible with the 
InterTEL board (section \ref{sec:intertel}).

Between 2016 and 2018 the L0 trigger primitive generation for the RICH, which has 2000 TDC channels distributed over 4 TEL62s, did not use 
inter-connected boards. Analogue sums of 8 channels available from the RICH front-end electronics were used, digitized in a fifth dedicated 
TEL62 board.

In the PPs a preliminary clustering is performed. In the SL, clusters coming from the 4 PPs are merged, then used to generate the L0 trigger 
primitives.
The FPGA resources usage for the common logic (sec. \ref{sec:generic_fw}) plus this L0 trigger logic amounts to 75\% (45\%) of the available 
logic elements and 47\% (44\%) of the available memory for the PP- (SL-)FPGA.

A common 32-bit data format (``RICH format'') is used to optimise the design efficiency: each firmware module uses such format for input and 
output, allowing to easily add, delete or reuse modules within the project.
This format includes two kinds of paired 16-bit words, each identified by a 2-bit word~ID: cluster words and timestamp words. 
The choice of 16-bit words is made to match the bus width of the InterTEL board (section \ref{sec:intertel}).


Cluster words identify clusters of hits and contain: a 12-bit fine-time ($T$) representing the time of the cluster in units of 100~ps; 
a 8-bit hit multiplicity ($N$) indicating the number of hits of a cluster; a 8-bit cluster time sum ($S$), which is the sum of the time 
differences between the cluster's ``seed'' time and the times of individual hits belonging to the cluster. 
$S$ is used to efficiently compute the weighted average of the cluster hit times, and is a signed number whose value remains small even for 
clusters with a large number of hits.

Timestamp words contain a 28-bit word with 400~ns LSb, with the most significant digits of the global experiment time. There are 16 
timestamp values in each 6.4~$\mu$s time frame. 
In case there are no hits corresponding to a given timestamp, a fake cluster with $N=0$, $S=0$ and the maximum possible $T$ is generated. 
Such clusters are called \emph{speed-data} and are used to control the firmware latency, while they are ignored by all modules.

\subsubsection{Overall L0 trigger logic}
\begin{figure}[hbt]
  \centering
  \includegraphics[width=\linewidth, clip]{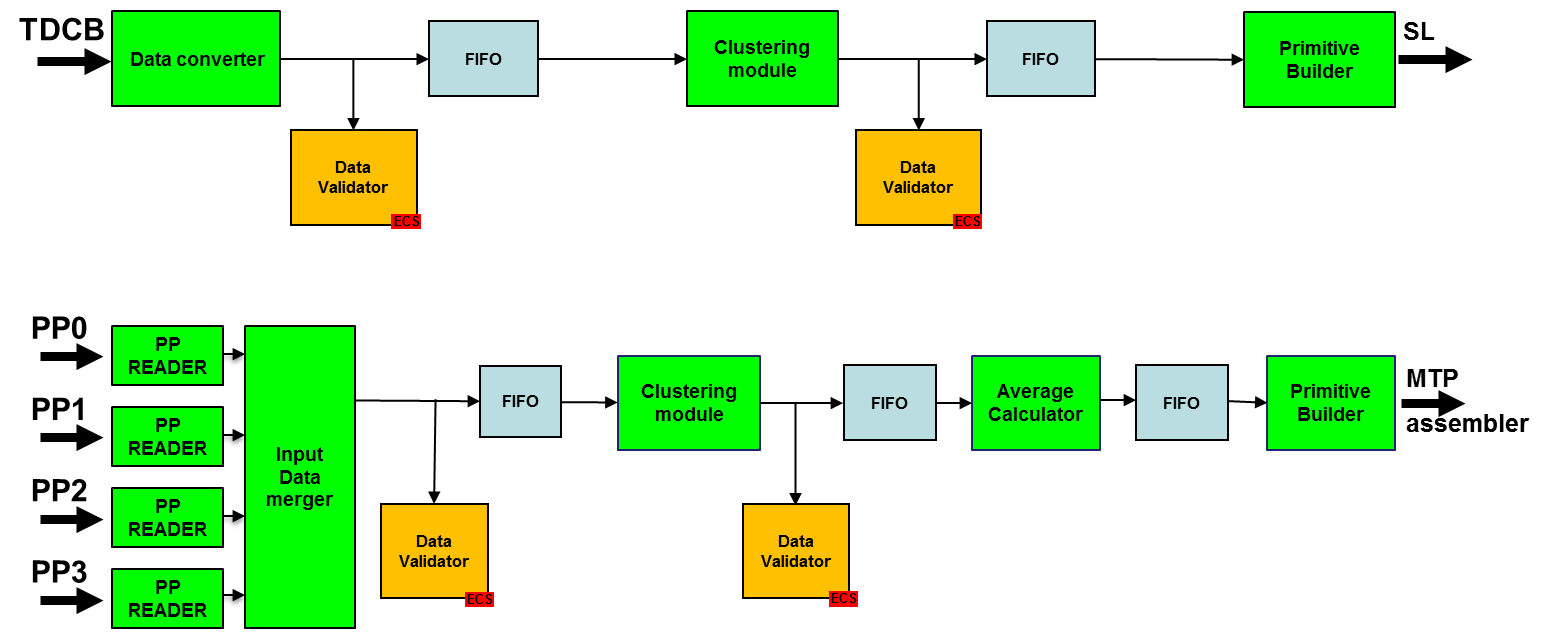}
  \caption{PP- (top) and SL-FPGA (bottom) RICH/NA48-CHOD firmware block diagrams.}
  \label{fig:RICH_PPandSL}
\end{figure}

Fig.~\ref{fig:RICH_PPandSL} shows the block diagrams for PP- and SL-FPGA firmware; the clock frequency is 160~MHz for all modules.
Primitives produced by the firmware represent time clusters of hits belonging to the same event.

In the initial stage of the PP-FPGA firmware the TDCB hits are formatted into clusters with $N=1$, $S=0$, and $T$ is set as the hit time. 
Such clusters are time-sorted at the 25~ns level, i.e. using the 4 most significant bits of $T$.
Data Validator (DV) modules check the consistency of the data format throughout the firmware: all timestamps and cluster data words must be 
sorted at the 25~ns level, and after a timestamp there must be at least one data word. If an error occurs a flag is set, and it has to be 
delivered in the EOB packet.
The following Clustering Module (CM), discussed in detail in section \ref{sec:l0clusteringmodule}, merges clusters, which are then sent to the 
SL-FPGA by a Primitive Builder (PB) module.

In the SL-FPGA, the Data Merger (DM) module merges the clusters coming from the 4 PP-FPGAs and potentially of the InterTEL boards. 
It reads data at a rate of 1 word per clock cycle, preserving the 25~ns time ordering and skipping timestamps and \emph{speed-data}. 
The module consists of 3 identical sub-modules organized in a two-level tree structure: each of the two branches merges the data from two of 
the PP-FPGAs, while the root merges the data from the branches. When using InterTEL boards, a 3-level DM structure can be implemented. 
Each module is purely combinatorial, with a FIFO buffer both on the input and on the output. Special care was taken to reduce the length of 
the paths and the number of combinatorial levels in each module.
DM sub-modules start working only if both input FIFOs are non-empty. This could lead to starvation of data in the FIFOs, \emph{e.g.} when one 
PP-FPGA receives hits from TDCBs faster than others. \emph{Speed-data} words are used to avoid starvation: since each PP-FPGA produces at least 
16 clusters per time frame, FIFOs cannot be empty for long periods of time. The $T$ value of \emph{speed-data} is set to the maximum possible 
value in order to give priority to real data in the DM.

After another Clustering Module, the Average Calculator (AC) module computes the weighted mean of the clusters' times: $T$ is increased by the 
$S/N$ ratio and $S$ is then reset to zero. The $S/N$ ratio is computed with a 8-bit FPGA-embedded divider.
Finally the SL-FPGA Primitive Builder (PB) formats the incoming data in the standard NA62 MTP format, computes the ID for each primitive and 
sends it to the standard MTP assembler module.

\subsubsection{L0 clustering module}
\label{sec:l0clusteringmodule}
\begin{figure}[htbp]
  \centering
  \includegraphics[width=0.9\linewidth, clip]{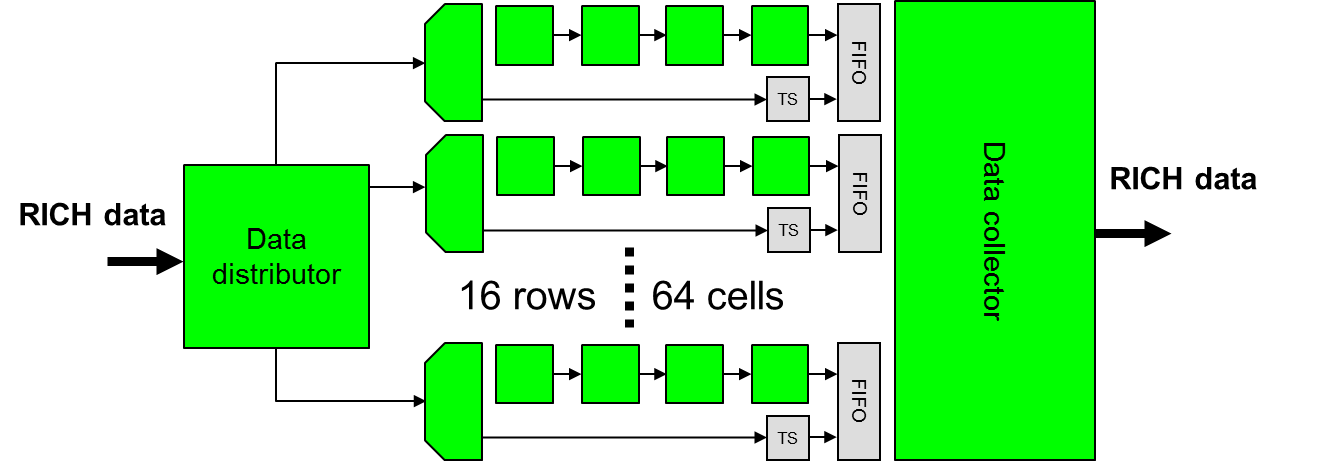}
  \caption{Block diagram for the RICH/NA48-CHOD L0 clustering module.}
  \label{fig:RICH_clustering}
\end{figure}

The CM consists of sub-modules, as shown in fig.~\ref{fig:RICH_clustering}. It merges clusters of hits or of other clusters that are closer 
in time than a programmable value. The comparison is performed only on $T$, while $S$ and $N$ are only taken into account after the comparison 
of $T$ has been made.

The Data Distributor (DD) module feeds the incoming data into so-called clustering rows. There are 16 rows, each made of 4 cells and a 
time slot (TS) register. Each row elaborates input hits belonging to a specific 25~ns time slot. By construction, a row can 
handle up to 4 clusters per 25~ns time slot. Any cluster beyond the fourth is discarded, and an error flag is set, to be sent in the EOB 
packet. Each cell consists of a comparison module, an embedded 18-bit multiplier and a correction module, all interconnected by FIFOs. 

The DD rearranges data into fine-time and TS, sending the former to the cells and the latter to the TS register of the proper 25~ns row. 
The DD is designed to send each hit into one of two adjacent rows (rows 15 and 0 are considered adjacent), depending on the time of the hit, 
and hits belonging to the same event which happen to be split over two adjacent 25~ns time slots can be merged together. All the computations 
within the cell are done on 9 bits: 8 bits for $T$ and 1 more bit to handle adjacent 25~ns time slots.

The comparison of the $T$ values and the cluster merging are performed by the cells: each cell initially stores the first received hit or 
cluster as a seed and defines a new cluster with $T_s$, $N_s$ and $S_s$ values in input.
If the $T_i$ value of an incoming 
cluster matches the stored $T_s$ within a programmable time window, $N_s$ and $S_s$ are updated: $N_s = N_s + N_i$, $S_s = S_s + (T_i - T_s) N_i$.
If $T_i$ does not match $T_s$, the incoming cluster is passed to the adjacent cell, which performs the same operations. If such cluster has a 
$T_i$ that is greater than $T_s$, the receiving cell increases an internal time-position field that will be used to time-order the clusters. 

The \(n\textsuperscript{th}\) row can be read out when the \((n+2)\textsuperscript{th}\) row is being filled: this is done by \emph{flushing} 
the row. Cells act as a shift register, writing the time position register and the $T_s$, $N_s$ and $S_s$ of the stored cluster to the row output 
FIFO.
In order to keep the firmware data throughput at 1 word per clock cycle, the flushing of a row must be completed before the row needs to be 
filled again. The output latency $L_o$ of a cluster is given by \(L_o = 2d+f+m\), where $d$ is the number of cells per row, $f$ is the delay 
of the output FIFO and $m$ is the latency of the cell multiplier. In our case, \(L_o = 2\cdot4+3+3=14\). The number of rows must be greater 
than $L_o$, and was set to 16 so that there will always be a free row to fill.

Finally, the Data Collector (DC) module retrieves data from the rows' output FIFO and converts them into the RICH format. The DC module is 
divided into four parts. The first one reads the data from the rows, being able to switch between rows without missing a clock cycle.
The second part sorts the clusters of a row by addressing RAM blocks with the time position field of the clusters. 
After the sorter, clusters with hit multiplicity outside a programmable range are discarded: this allows to reduce the noise related to 
events with too low or too high hit multiplicity. The fourth part of the DC reads the sorted remaining clusters and converts them to the 
RICH format.

\subsubsection{Test and performance}
\begin{figure}[tbp]
  \centering
  \includegraphics[width=0.8\linewidth, clip]{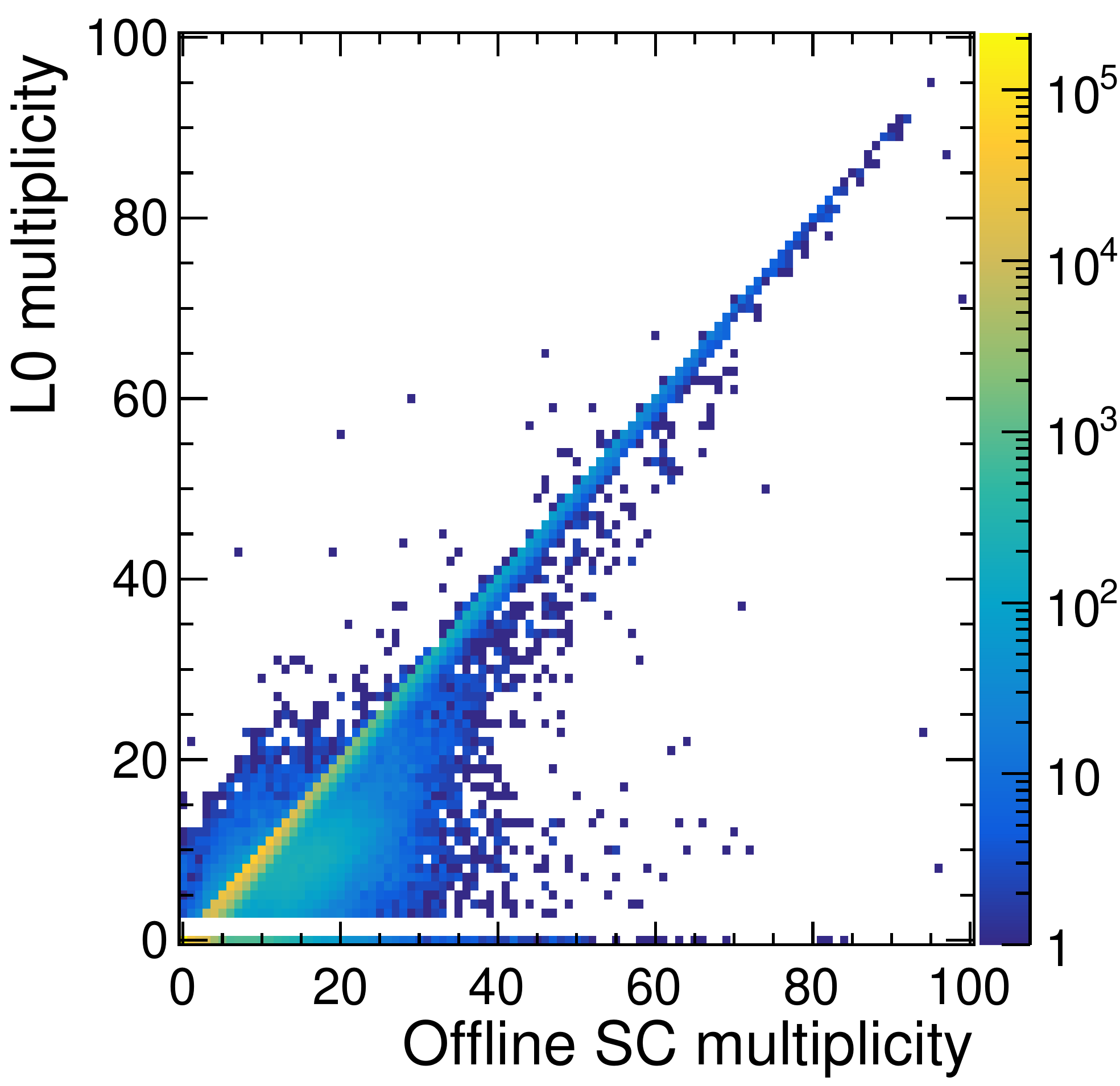}
  \caption{Correlation between the hit multiplicity computed by the RICH L0 firmware and in off-line data analysis.}
  \label{fig:RICH_correlation}
\end{figure}

The design was simulated and implemented in the actual system since the beginning of the 2016 physics run. Data triggered by other sub-detectors 
were analysed and compared to the output of the RICH firmware, in order to check its behaviour. The analysis isolates the events in the RICH 
detector, computing their time and hit multiplicity. Fig.~\ref{fig:RICH_correlation} shows the high correlation between the hit multiplicity 
computed in the off-line analysis and by the on-line RICH L0 firmware. The module inefficiency is computed by considering all events that have 
zero L0 hit multiplicity but non-zero analysis multiplicity, and is measured to be 1.24\%. The probability of false positives is computed by 
considering all events with zero hit multiplicity in the off-line analysis and non-zero L0 multiplicity, and is 0.005\%.

\begin{figure}[thbp]
  \centering
  \includegraphics[width=0.9\linewidth]{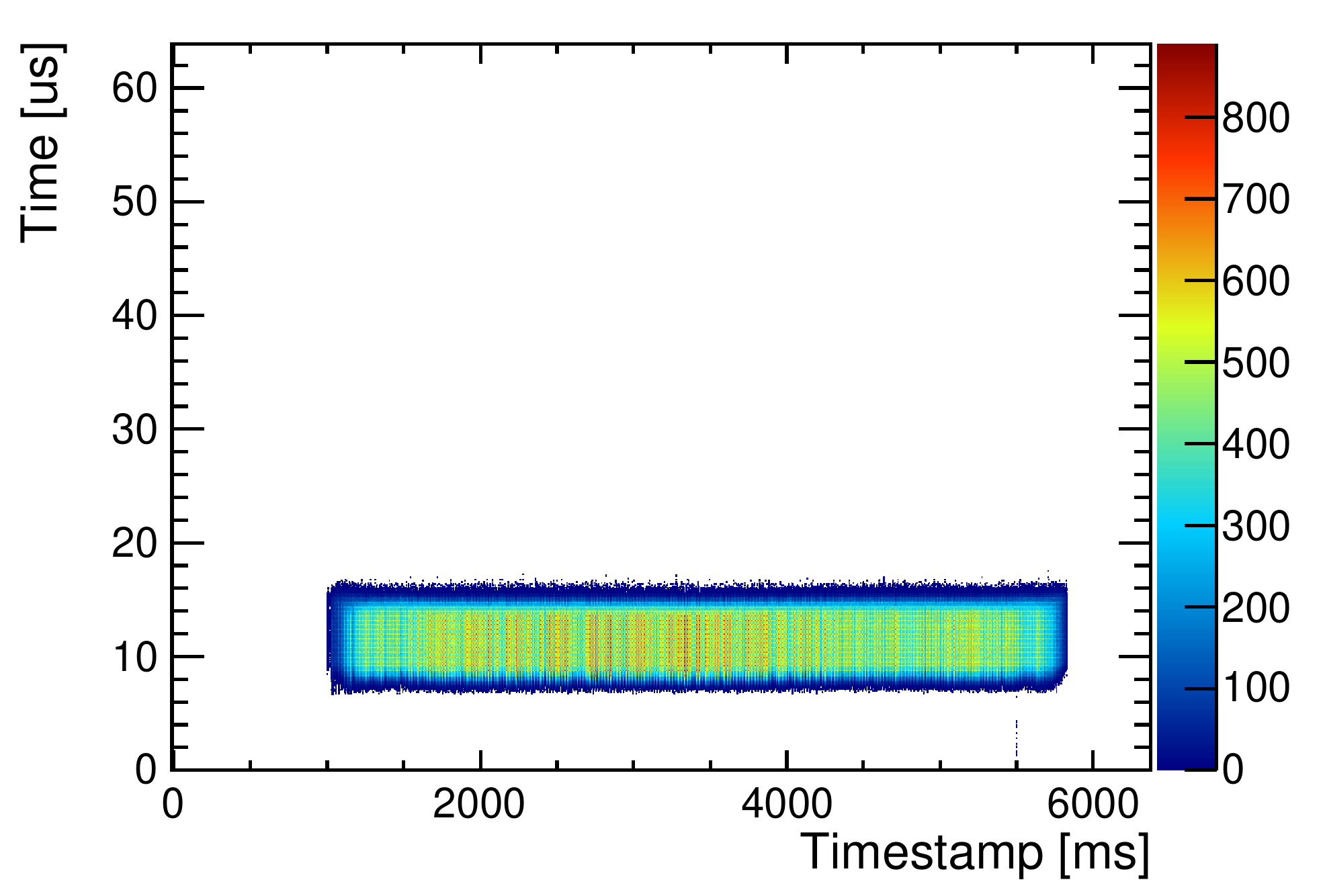}
  \includegraphics[width=0.9\linewidth]{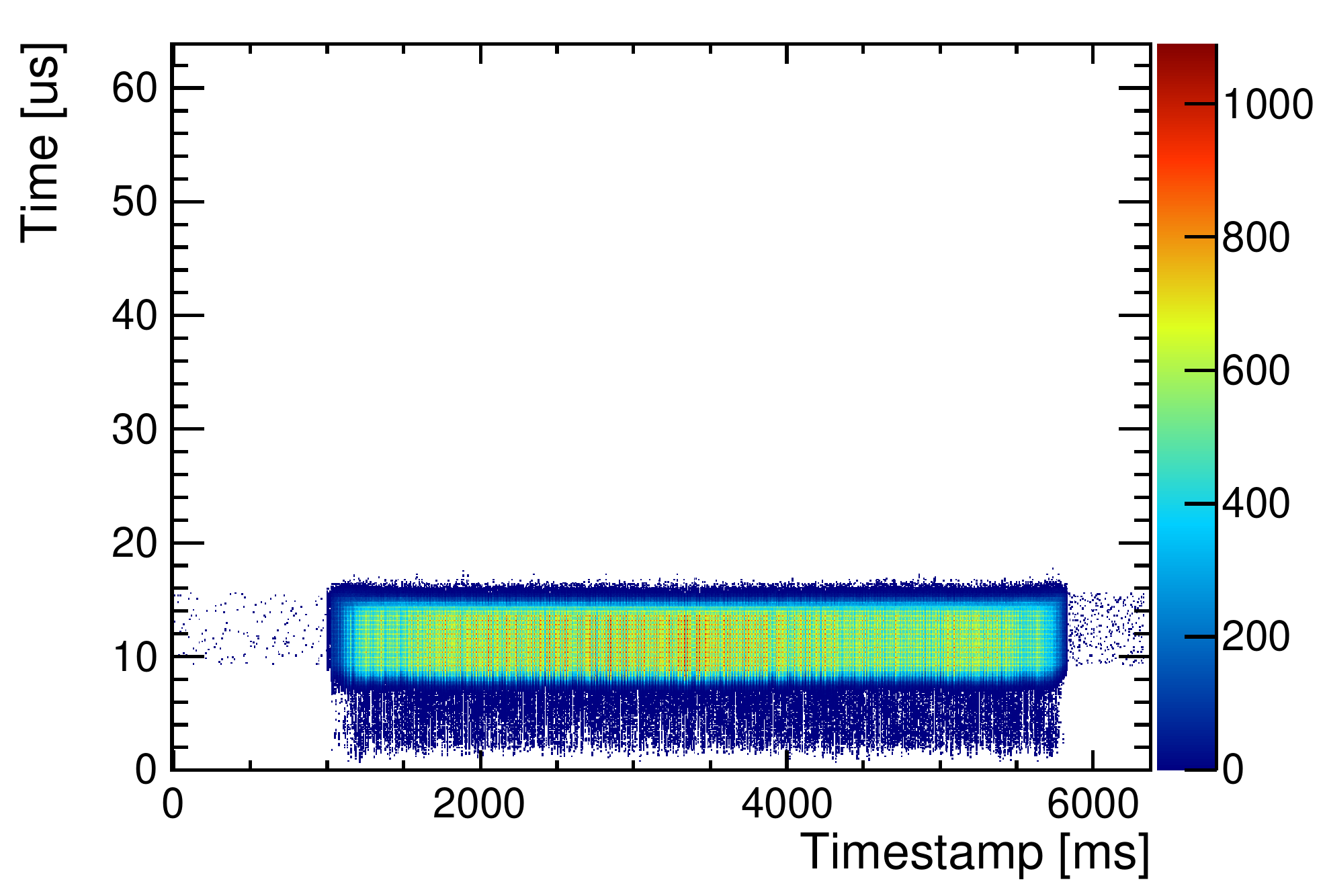}
  \caption{Primitive generation delay for the RICH (top) and NA48-CHOD (bottom) sub-detectors.}
  \label{fig:delayRICH}
\end{figure}

The top plot in fig.~\ref{fig:delayRICH} displays the 25~ns time slot contained in a L0 trigger primitive versus the generation time of the 
MTP including such primitive.
It can be seen that the firmware latency is very stable, being between 2 and 3 (6.4~$\mu$s-long) time frames for the RICH sub-detector. 
The same information for the NA48-CHOD detector is shown in the bottom plot: the latency is lower than in the RICH case, and varies 
between 1 and 2 time frames, the reason being the 15\% higher hit rate in the NA48-CHOD, which reduces the latency of the DM and the DDs.

\subsection{LAV L0 trigger}
The Large Angle Veto (LAV) L0 primitive generator firmware \cite{LAV-fw} works under the assumption that input digital signals originate from 
the LAV front-end boards \cite{LAV-frontend}.
These boards are double-threshold discriminators producing two digital LVDS output signals from each analogue input signal.
Analogue input channels will be referred to as “blocks” in the following, in analogy with the LAV lead-glass blocks.
In 2016-2018 the LAV L0 trigger primitive generator only used information from the 12th LAV station.
If information from more LAV stations were to be used, the relevant TEL62 boards could be connected together using InterTELs (section 
\ref{sec:intertel}).

The LAV L0 primitive generator starts by associating the low- and high-threshold crossings in a specific block to build hits, and performs a slewing correction on them.
The block hits are then clustered together, based on a programmable time interval.
A LAV L0 trigger primitive is built from each cluster.
%
The FPGA resource usage for the common logic together with the LAV L0 trigger logic amounts to 79\% (28\%) of the available logic elements 
and 51\% (39\%) of the available memory for the PP- (SL-)FPGA.

\subsubsection{PP-FPGA logic}
\begin{figure}[hbt]
  \centering
  \includegraphics[width=1.0\linewidth]{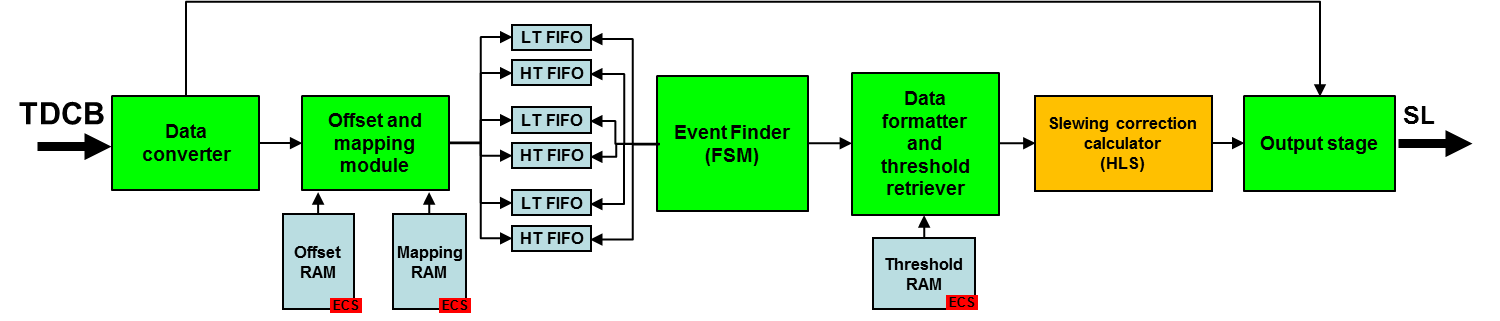}
  \caption{Block diagram of LAV L0 firmware in the PP-FPGA.}
  \label{fig:LAV_PP}
\end{figure}

Fig.~\ref{fig:LAV_PP} shows a scheme of the LAV L0 firmware in the PP-FPGA, whose main task is to generate a \emph{block hit}, if both high and 
low thresholds are crossed for a block within a programmable time interval.
After the times of such hits are corrected for slewing, they are sent to the SL-FPGA.

The LAV firmware reads data produced by the HPTDCs through a FIFO buffer. The input stage reads its content at the rate of 1 word per clock 
cycle (160~MHz) and sends it to the next firmware module. If the data word is an end-of-frame counter, a global end-of-frame signal is sent to 
the Event Finder module. 

The time-offset and channel mapping module receives data from the input stage continuously. It checks if a word refers to a leading-edge or a 
trailing-edge time measurement and, if this is the case, it retrieves the proper offset value from the time-offset RAM and the remapped 
channel number from the mapping RAM and applies them to the current data word. Such memories are fully programmable by the user at configuration 
time through ECS.
The total input-output latency of this module is 4 clock cycles.

The block time is reconstructed by associating the High Threshold ($HT$) and Low Threshold ($LT$) crossing times. For this purpose data are 
separated into 128 FIFO buffers depending on the channel number by the Channel-Selector module, redirecting data words to the corresponding 
channel FIFO.
Frame timestamp words are sent to all FIFOs in parallel, while leading times are sent only to the proper channel FIFO and trailing times are 
discarded.
At this stage the word size is reduced from 32 to 22 bits, by using a single bit as word type flag (timestamp or leading-edge fine-time), 
discarding the time bits overlapping between frame timestamp and pulse time, and removing the channel number which is encoded in the FIFO index.

The FIFOs are arranged in 64 blocks, each composed of a \emph{high-threshold FIFO} (HF) and a \emph{low-threshold FIFO} (LF), 16 and 
32 words deep respectively, the difference being due to the different rates expected.
Each FIFO corresponds to two registers: a 18-bit fine-time register and a 22-bit coarse time register. When a FIFO is not empty, data is read 
out and stored in these registers, forming a 40-bit output word.
If both the HF and LF contain a complete word, \emph{i.e.} both the fine-time and coarse time registers are full, then the block is ``ready'' 
to be read-out by the next firmware module.

The Event Finder (EF) module is a FSM which looks for ``ready'' blocks and builds a block hit. When a block is ``ready'', the module reads its 
$HT$ time and stores it, then it reads the $LT$ time and subtracts it from the stored $HT$ time. If the resulting difference matches 
programmable limits, a block hit is generated, otherwise the earlier of $HT$ and $LT$ is discarded.
For each generated block hit the EF produces an output data word with block ID (6 bit), time since the start of the burst (40 bit) and the 
difference of the HT and LT crossing times, also called rise time (8 bit).


In order to perform a slewing correction, the two threshold values, $V_{\mathrm{low}}$ and $V_{\mathrm{high}}$ set in the LAV front-end boards 
for every channel, are required. Such values are stored via ECS into the Threshold RAM, with 128 12-bit locations, where the least-significant 
bit corresponds to 0.1 mV and the maximum value is 409.5 mV. The Slewing Correction module retrieves the appropriate threshold values from the 
RAM and evaluates  the corrected time $t$ by linearly extrapolating the starting point of the analogue signal, based on the measured 
threshold-crossing times $t_{\mathrm{low}}$ and $t_{\mathrm{low}}$. This module is implemented using High Level Synthesis via Catapult\reg \,
\cite{Catapult}. 


\subsubsection{SL-FPGA logic}

\begin{figure}[hbt]
  \centering
  \includegraphics[width=1.0\linewidth]{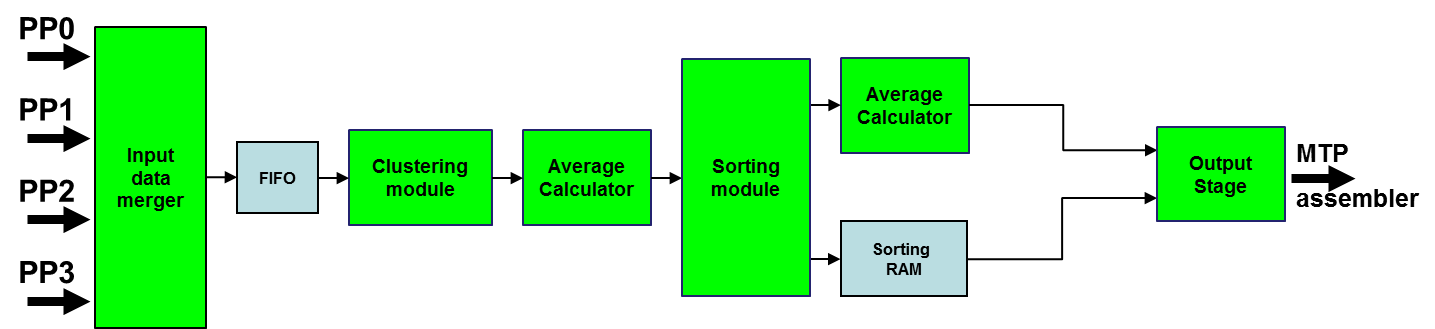}
  \caption{Block diagram of the LAV L0 firmware in the SL-FPGA.}
  \label{fig:LAV_SL}
\end{figure}

Fig.~\ref{fig:LAV_SL} illustrates the LAV L0 primitive generating firmware in the SL-FPGA. The firmware collects the block-hits coming from the 
PP-FPGAs and creates clusters of them according to a programmable time interval. The time of each cluster is calculated by averaging the times 
of the block hits contained in the cluster. Then, a LAV L0 trigger primitive is generated from each cluster. Finally, the primitives are 
time-sorted and sent to the MTP assembler module.

The first stage is the Data Merger (DM) module, a FSM handling the 4 input FIFOs containing data coming from each PP-FPGA.
The DM reads data from the first non-empty FIFO of all enabled PP-FPGAs; in case two or more FIFOs are not empty at the same time, the 
priority switches cyclically.
The DM produces a global End-of-Frame (EoF) word when it receives an EoF word from each of the enabled PP-FPGAs.
Output data from the DM is stored in another FIFO buffer.

The Clustering Module (CM) reads data from such FIFO and creates clusters from block hit times that match within a programmable interval.
At this stage, the information about block number and rise time is discarded.
The CM is composed of 32 cells, connected in series in a shift-register fashion. 
The first time value received by a cell is stored, creating a cluster. When a cell receives further time values, it evaluates their difference 
with respect to the stored value and compares it with a programmable limit: if the difference is within the limit, the cell adds it to the 
stored cluster, otherwise the cell sends the new time value to the next cell.
The limits of the matching interval are asymmetrically programmable (up to $\pm$12.5~ns) through ECS.
When the CM reads the global EoF word, the whole module acts as a shift-register, outputting the register contents for each cluster.
Such information is sent to the Average Calculator (AC) module, which computes the average of hit time differences with respect to the first 
block hit time. Such value is finally added to the stored time, obtaining the final fine-time for the cluster.
This procedure is used in order to reduce the size of the words used in the division, and therefore save FPGA resources.

The Sorting Module (SM) receives cluster data from the AC and time orders them. The architecture of the SM is similar to that of the CM: 
it is composed of 32 basic cells connected in series.
Each cell receives a word containing the cluster time and number of hit values; the first is stored, and a time position register initialised at 
zero. When subsequent data arrive, if their time value is greater than the stored one, the cell just sends such data to the next cell.
On the contrary, if the time is smaller than the stored one, the cell increments the value of its time position register before sending the 
data to the next cell.
When the EoF signal is received, the whole module acts as a shift-register outputting all the cluster times, number of hits, and 
respective time position values.
Such data is stored into a RAM addressed with the time position value, and the number of clusters is also counted.
Finally the RAM is read out starting from address 0 up to the number of clusters previously counted, producing time-sorted L0 trigger 
primitives. 

\subsubsection{Test and performance}
\begin{figure}[thbp]
  \centering
  \includegraphics[width=0.9\linewidth]{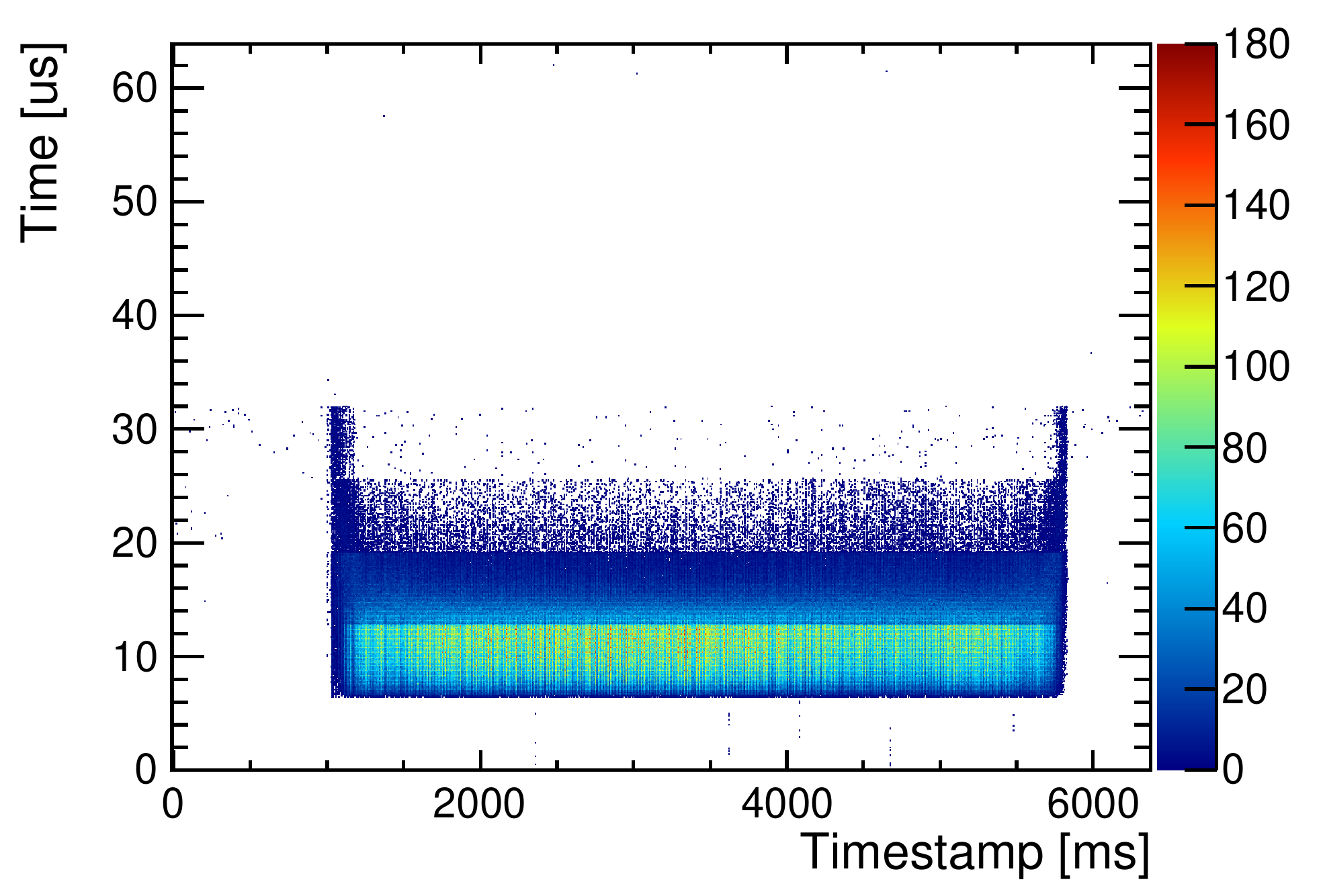}
  \caption{Primitive generation delay for LAV12.}
  \label{fig:delayLAV}
\end{figure}

The LAV L0 trigger generating firmware design was simulated and implemented in the TEL62 board for the most downstream LAV station (LAV12) 
and tested in the NA62 2015 run. It was fully operational and was used in veto on the $K^+ \rightarrow \pi^+ \nu \overline{\nu}$ trigger 
stream during the 2016-2018 data-taking.
The firmware performed according to specifications, producing an average rate of $\sim$ 1~MHz of primitives with an average beam intensity
of $18 \cdot 10^{11}$ protons on target per burst.
Fig.~\ref{fig:delayLAV} shows the 25~ns time slot contained in a trigger primitive versus the generation time of the MTP in which such 
primitive is contained: the firmware latency is between 1 and 4 (6.4~$\mu$s-long) time frames. 

\subsection{MUV3 and CHOD L0 trigger}

The MUV3 and CHOD detectors are both charged-particle detectors. Each is a single plane of plastic scintillator, segmented in the plane 
orthogonal to the beam to form differently-sized ``tiles''. Similarities among the two sub-detectors led to the development of a common L0 trigger 
logic, with slight sub-detector specific differences.

The MUV3 detector is composed of 148 tiles, with 140 ``outer'' tiles, which can be uniquely assigned to one sub-detector ``quadrant'', and 8 
``inner'' tiles, closer to the beam pipe where the rate is higher, which cannot be assigned to any quadrant.

About 14~MHz of muons are expected to cross MUV3, half of that rate being in the inner tiles, with more than 3~MHz in one of the tiles.
Each tile is read-out by two photo-multipliers 20~cm downstream, with their glass window facing upstream. About 8\% of incident muons 
pass through the glass window of one of the two PMs of a tile, causing \v{C}erenkov light to be produced there; the signal from such light 
reaches the TDC about 2~ns earlier than the signal due to scintillation light.

The CHOD detector is composed of 152 tiles, arranged into four identical ``quadrants'', each read-out by two silicon PMs (SiPMs) situated outside 
of the detector acceptance and connected to the tiles by wavelength-shifting fibres.
About 10~MHz of charged particles are expected to cross the CHOD, with a more uniform distribution among tiles than in the case of MUV3.

Each PM of the above sub-detectors is connected to one channel of a Constant Fraction Discriminator, which outputs fixed-length digital pulses 
to the TDCs. As the pulse has a fixed length, the trailing-edge measurement made by the TDC does not contain useful information, and is 
discarded. Higher-rate PMs are assigned to channels with the highest readout priority within the HPTDC (channel 0 of each 8-channel group).
The MUV3 is equipped with 3 TDCBs (12 HPTDCs) to accommodate the 296 PMs, with two 8-channel groups being uniquely reserved to the 
highest-rate tile; CHOD uses 3 TDCBs, one of those being shared with another sub-detector, so that 10 HPTDCs accommodate its 304 SiPMs.

In both detectors, so-called candidates are formed by combining the hits in the two PMs of each tile; hits in close time coincidence in both 
PMs result in a ``tight'' candidate. 
To counteract the time jitter related to early \v{C}erenkov light, the latest of the two close-in-time hits defines the candidate time in 
MUV3, while the average time of the two hits defines the CHOD candidate time.
A PM hit that is not in coincidence with a hit in the other PM of the same tile forms a ``loose'' candidate, for which the candidate time 
is simply the hit time. 
Both tight and loose candidates are produced to maintain high efficiency in case one of the two channels of a tile is not working properly, 
or if the time alignment between the two channels is poor.
Typical coincidence windows are 10~ns for MUV3 and 15~ns for CHOD (reflecting the intrinsic time resolutions).



MUV3 is used in the L0 trigger to tag muons, while the CHOD is used to tag any charged particle.
The first part of the L0 trigger logic, implemented in the PP-FPGA, builds candidates based on hits from the two PMs in each tile. The second 
part, implemented in the SL-FPGA, merges coincident candidates into L0 trigger primitives, in such a way that the primitives are produced in 
time order. The L0 trigger primitives encode trigger conditions, which are used to make L0 trigger decisions in the L0TP. 
For the CHOD those conditions are: at least one candidate in any quadrant (Q1); less than 5 tight candidates (UTMC); candidates in at least two 
quadrants (Q2); and candidates in at least one pair of diagonally-opposite quadrants (QX). 
For the MUV3 the conditions are: at least one candidate (M1); at least one candidate in an outer tile (MO1); at least two candidates in outer 
tiles (MO2); and candidates in at least one pair of diagonally-opposite quadrants, recalling that inner tiles are not assigned to a quadrant 
(MOQX). The MUV3 conditions that utilise only the outer tiles are used to select muons; using only the outer tiles substantially reduces the 
trigger rate while only slightly reducing the trigger acceptance.

\subsubsection{PP-FPGA trigger logic}
\begin{figure}[hbt]
  \centering
  \includegraphics[width=0.8\columnwidth]{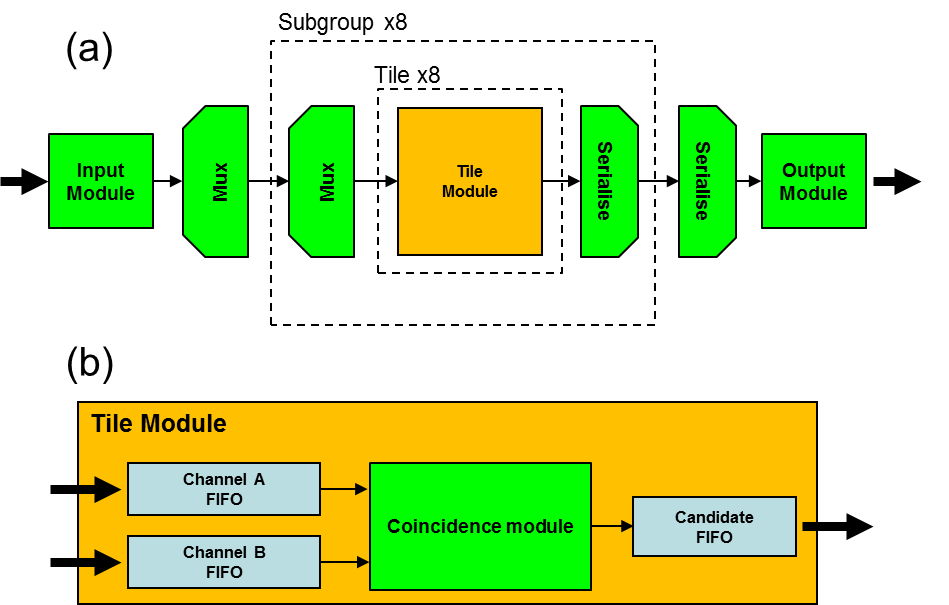}
  \caption{Schematic of the MUV3/CHOD PP-FPGA L0 trigger firmware (a), with detailed schematic of the tile module (b). }
  \label{fig:CHOD_PP}
\end{figure}

Fig.~\ref{fig:CHOD_PP} shows the outline of the PP-FPGA logic.
TDC hits are read from the PP-FPGA TIB, arranged in 6.4~$\mu$s frames and time ordered at the level of 25~ns. Combining the frame timestamp 
and the hit fine-time, an absolute time since the beginning of the burst can be associated to each hit.

The logic requires hits from the two PMs of the same tile to be in adjacent TDC channels: to enforce this, hit channel IDs are remapped, 
based on the contents of a programmable memory initialized by the user.
As the CHOD detector shares one TDCB with another sub-detector, hits from the other detector are removed in the first firmware stage, using 
a flag in the above channel-mapping memory which can be also used to mask malfunctioning channels.

TDC hits are then sent to one of 64 ``tile'' modules according to their (modified) channel ID, via two stages of multiplexer modules, first 
separating into 8 subgroups of 16 channels each and then separating each subgroup into 8 tiles of two adjacent channels.
Each tile module contains two ``channel FIFO'' buffers used to store hits from the two channels of the same tile, plus a coincidence module 
that compares absolute hit times.
If hits have times within the (programmable) coincidence window they are combined into a tight ``candidate'', otherwise the earlier hit is 
converted to a loose candidate, while the later one remains in the buffer. All candidates are written to a ``candidate FIFO'' buffer. 

In normal operating mode, hits are converted to loose candidates if they remain in the channel buffer longer than a time-out of 125~ns; 
this feature prevents hits from building up in the channel buffers in case one channel has more hits than the other in the tile. 
However, once all the hits of a frame have reached the channel buffer, the firmware enters ``loose mode'': the time-out is reduced to 12.5~ns, 
so that remaining hits can be processed before the arrival of the next frame, without affecting the production of tight candidates.
If the next frame arrives before all hits have been converted into candidates, all buffers are cleared to make way for the incoming data.

The 8 candidate buffers of each subgroup are serialised, with priority given to larger channel IDs, and written to a subgroup output buffer; 
such buffers are serialised in the same way, and the candidates sent to the output module.

The output module writes the information of each candidate in a 32-bit data word, stored in an output buffer FIFO, for transfer to the SL-FPGA,
again in a timestamped 6.4~$\mu$s frame format. 
Each data word records the candidate time within the frame, its tile number, flags to indicate it as as an inner or outer tile, and the 
quadrant number. The latter two pieces of information are extracted from a dedicated programmable memory.
A frame ends when all the hits in a frame have been processed or when the next frame has arrived.

\subsubsection{SL-FPGA trigger logic}
\begin{figure}[hbt]
  \centering
  \includegraphics[width=0.8\columnwidth]{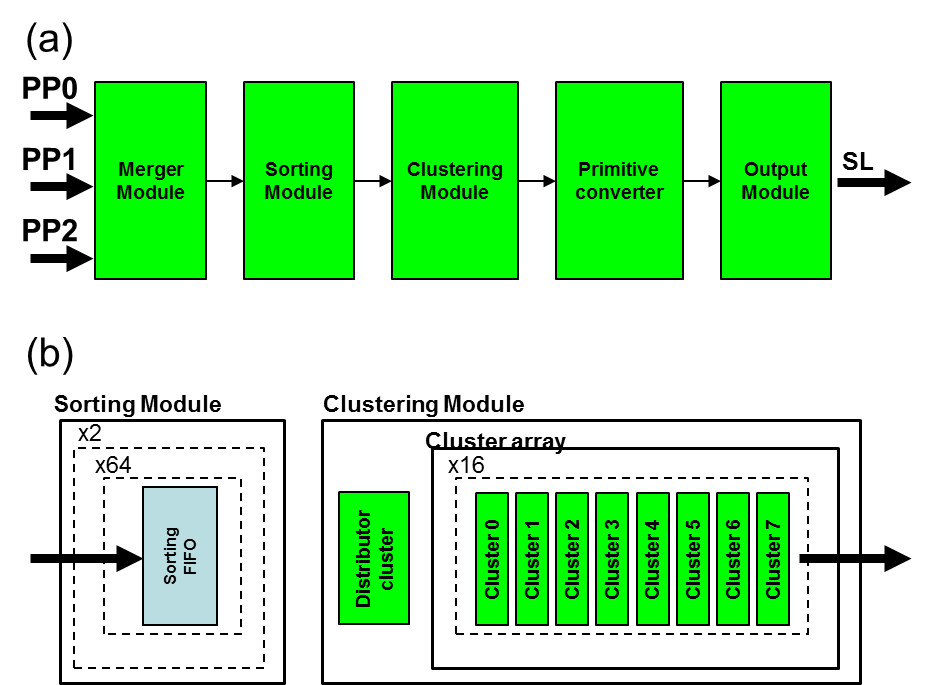}
  \caption{Schematic of the MUV3/CHOD SL-FPGA L0 trigger firmware (a), with detailed schematic of the clustering module (b). }
  \label{fig:CHOD_SL}
\end{figure}

The CHOD/MUV3 L0 trigger logic in the SL-FPGA is shown schematically in fig.~\ref{fig:CHOD_SL}.
The SL-FPGA receives data from 3 PP-FPGAs. The data are read from input FIFO buffers using a round-robin technique whenever one of the buffers 
is not empty.
The frame timestamp is combined with the time information in the data words to recreate the candidates' times. 
Reading from a buffer stops at the end of each frame, until all other buffers have also completed reading and each candidate has been sent to 
the sorting module; this maintains data synchronisation from the three input buffers.

The sorting module contains two sets of 64 ``sorting FIFO'' buffers (fig.~\ref{fig:CHOD_SL}).
Candidates in consecutive frames are sent alternately to the first or second set of buffers, and they are divided among the 64 buffers in each 
set according to their time within the frame, each buffer storing hits related to a 100~ns time interval, with earliest hits in the first and 
latest hits in the last buffer.
Once all candidates from a frame have been written, they are passed to the clustering module.

Each sorting buffer is read first-to-last, so that candidates are read in time order (at the 100~ns level).
The first candidate in each frame is converted to a ``cluster'' format, and is set as the ``distributor cluster''.
The cluster format contains: the cluster seed time (T), being the time of the candidate that created the cluster; the summed time-difference 
(DT) between any merged candidates and the cluster seed time; the number of candidates (N) merged into the cluster; two counters to record the 
number of tight and loose candidates; two counters to record the number of inner/outer tile candidates; and four flags to record which quadrants 
the merged candidates are assigned to.

The second candidate in the frame is read and the time of the candidate is compared to the seed time of the distributor cluster.
If the difference is less than a (programmable) matching time window, the candidate is merged into the distributor cluster. If not, the 
candidate is converted to the cluster format: if the candidate time was smaller (earlier) than the distributor cluster, the new cluster is sent 
directly to the cluster array; otherwise, the distributor cluster is sent to the cluster array and the new cluster becomes the distributor 
cluster.

The cluster array consists of 16 rows of 8 cells, each row corresponding to one of the sorting buffers. Each cell can store one cluster.
When a cluster is sent to an empty row, the cluster simply occupies the first cell in the row. 
The time of each following cluster is compared to the seed time of the cluster in the first cell of the row: if the difference is less than the 
(programmable) matching time window the two clusters are merged. If not, and the incoming cluster time is larger (later) than the existing 
cluster, the incoming cluster occupies the first cell and all existing clusters in the row are shifted to the next cell; otherwise, the 
incoming cluster just moves to the next cell. Typical matching time windows are 10~ns wide.


No more clusters are sent to the N$^{\mathrm{th}}$ row of the cluster array once the N$^{\mathrm{th}}$+2 row starts being filled.
Additionally, the last cluster written to the N$^{\mathrm{th}}$ row must be allowed time to move down the row eight times, corresponding to 
the eight cells in the row.
Once these two conditions are satisfied, the N$^{\mathrm{th}}$ row can be emptied.
The rows are emptied by reading the eight cells one-by-one, starting from the one with the largest index; this procedure ensures that clusters 
leave the sorting module in time order.



Clusters are converted to trigger primitives in the primitive converter module.
Trigger primitives consist of an absolute primitive time $PT = T + DT/N$ (recall that $T$, $DT$ and $N$ are stored in each cluster), and a 
primitive ID encoding up to 16 sub-detector conditions, computed based on the different counters and flags recorded in the cluster.
The primitive ID is sub-detector specific: the CHOD contains flags for the Q1, Q2, QX, and UTMC conditions; while the MUV3 contains flags for 
the M1, MO1, MO2, and MOQX conditions.

\subsubsection{Test and performance}
Several checks are performed, both on-line and off-line to validate and monitor the correct functioning of the system, such as those 
concerning the correct assignment of tiles to quadrants, which is crucial for L0 trigger primitive generation.


\begin{figure}[ht!]
  \centering
  \includegraphics[width=0.9\columnwidth]{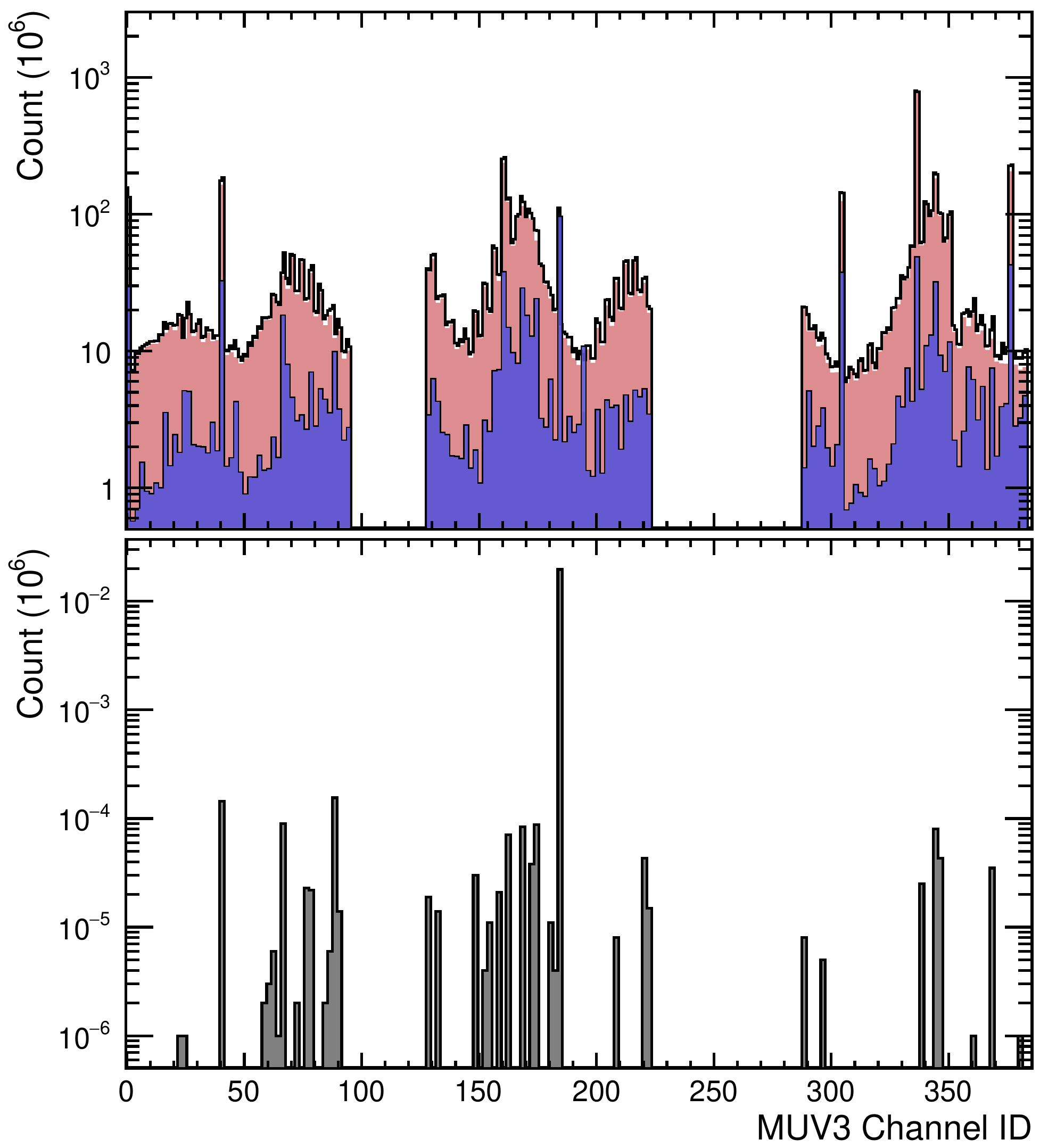}
  \caption{Top: Distribution of TDC hits in each MUV3 channel (black solid histogram), tight (light grey/mauve filled region) and loose 
  (dark grey/blue filled region) candidates in each tile. Bottom: computed number of missing MUV3 TDC hits per channel.}
  \label{fig:MUV3_missing}
\end{figure}

The total number of tight and loose candidates per tile in a burst, as well as the number of TDC channel hits, are written to the 
End-of-Burst data packet available off-line. Hence, any malfunctioning leading to hit losses can be monitored, by summing the number of TDC 
hits in the two channels of a tile and subtracting the number of loose candidates plus twice the number of tight ones: such checks show that 
hit losses are $\approx 2 \times 10^{-6}$ in the MUV3 logic and $\approx 4 \times 10^{-4}$ in the CHOD logic (fig.~\ref{fig:MUV3_missing}).
In the MUV3 logic the losses mostly appear in noisy channels and are likely due to anomalous TDC hits filling the channel buffers of the tile 
modules. In the CHOD logic the losses are likely due to the processing of TDC hits from one frame not being finished before the arrival of the 
next one; this effect is larger in CHOD due to the larger average hit multiplicity per event, further exacerbated by the presence of anomalous 
events in which tens of CHOD tiles are hit simultaneously.

\begin{figure}[ht!]
  \centering
  \includegraphics[width=0.9\columnwidth]{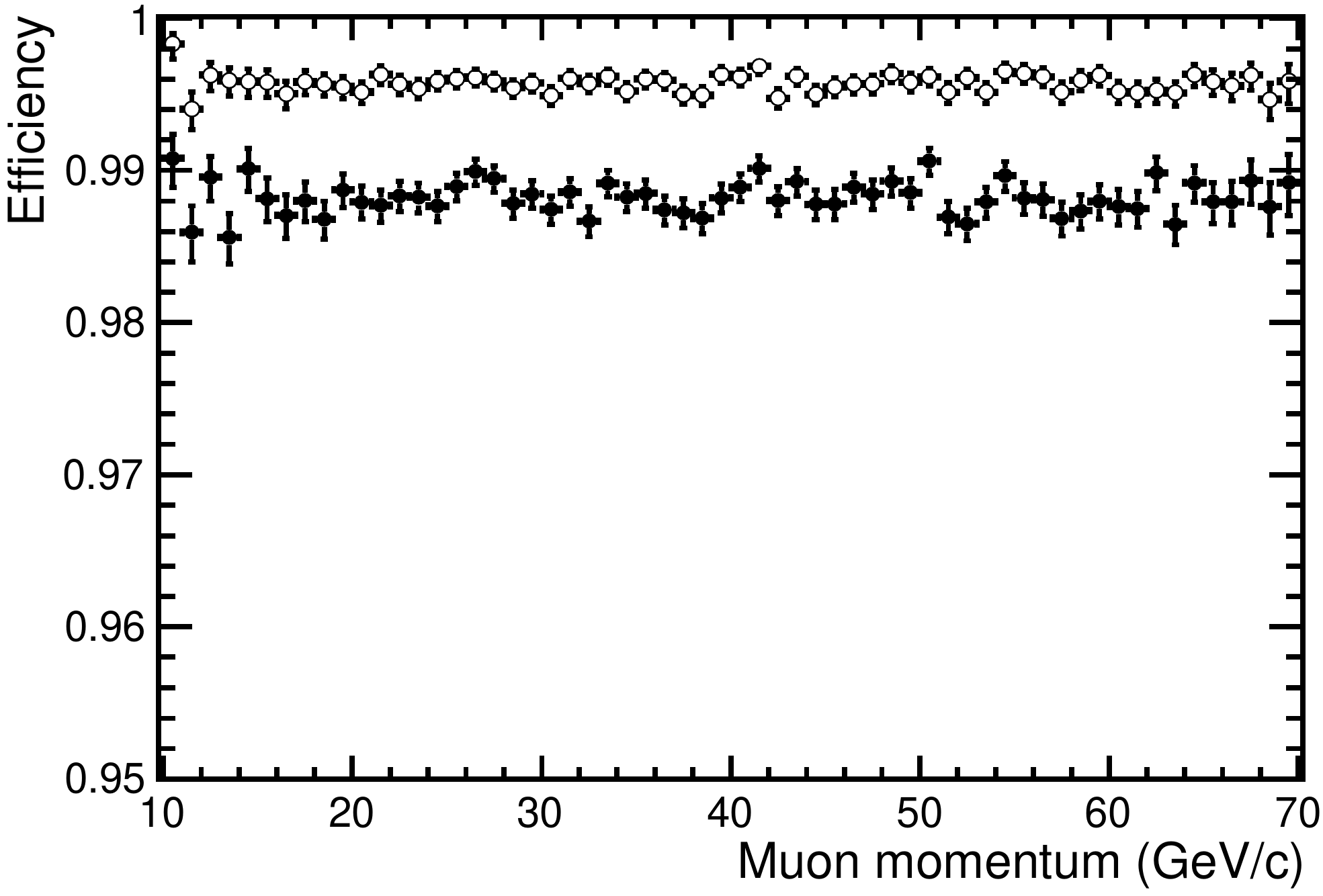}
  \caption{Efficiency of the CHOD Q1 (black dots) and MUV3 M1 (white dots) primitive trigger conditions as a function of track momentum.}
  \label{fig:MUV3_efficiency}
\end{figure}

The efficiency of MUV3 M1 condition is above 99.5\% while the efficiency of the CHOD Q1 condition is above 98.5\%, as measured using a sample 
of $K^{+} \to \mu^{+}\nu$ decays collected with a minimum-bias control trigger (fig.~\ref{fig:MUV3_efficiency}).
The FPGA resources usage for the common logic together with the MUV3 L0 trigger logic amounts to 74\% (62\%) of the available logic elements 
and 48\% (37\%) of the available memory for the PP- (SL-)FPGA. The figures for the CHOD trigger logic are the same, except for the PP-FPGA 
logic resources being 87\% due to the more complex time-averaging algorithm.

\section{ADC-based L0 trigger logic}
\label{sec:lkrl0}

Information from the NA62 calorimeters is combined in L0 trigger primitives that encode trigger conditions based on the number of coincident 
clusters and the total energy contained within those clusters.
To determine the number of clusters and the total energy, a cluster search is performed in the calorimeters.
In the LKr, the cluster search is performed in two steps, with two one-dimensional algorithms; the first step is implemented in the 
Front-End boards, while the second is implemented in the Merger boards.
In the other calorimetric sub-detectors the cluster search is performed in a single step, with a single one-dimensional algorithm implemented 
in the Front-End boards. 

In the case of the LKr, each Front-End board receives data from a vertical ``slice'' of the detector that is one super-cell wide and runs the 
full height of the LKr.
The cluster search in the Front-End boards comprises three steps. First, a peak search in space is made by seeking an energy peak in neighbouring 
super-cells at a fixed sample time. Next, a peak search in time is made by seeking an energy peak in neighbouring samples in each super-cell.
Finally, the energy in each peak is compared to a (programmable) threshold; peaks that contain less energy than the threshold are discarded.
For all remaining peaks a parabolic interpolation is performed, followed by a constant-fraction discrimination, to more-precisely determine 
the energy and time of the reconstructed peak.

The reconstructed peaks are propagated to the Concentrator board, where the peaks are merged into clusters and time-binned in a dual-ported 
circular RAM buffer.
L0 trigger primitives based on programmable conditions on energy sums and cluster multiplicities are produced at the output of the circular 
buffer. The L0 trigger primitives are then sent to the L0TP in UDP packets over the standard GbE links.

In the first NA62 data-taking period (2016-2018) only information from the LKr was utilised.
Further information on this trigger sub-system will be made available elsewhere.





\section{L0 trigger primitives readout}
\label{sec:primitive_readout}

An independent acquisition of the L0 trigger primitives was implemented, which allowed the collection and analysis of all generated L0 trigger 
primitives just a few seconds after the end of a burst.
In the early years of the experiment, this system was used extensively to commission the L0 trigger system.
During the data-taking, it was used primarily: to investigate any malfunction of the L0 trigger system; to monitor on-line the generation and 
synchronisation of the L0 trigger primitives; and to monitor on-line the instantaneous time profile of the beam.



The L0 primitive acquisition system is composed of 7 GbE switches\footnote{D-Link DGS-1100-08}, each dedicated to one of the L0 trigger 
primitive generating sources, and one rack-mounted acquisition PC with 7 dedicated GbE ports.
The switches are connected in a daisy-chain, forming a private network connected to the acquisition PC. The PC can fully configure the 
switches, allowing to enable or disable each primitive stream and the primitive acquisition independently.
Three ports in each switch are used for the L0 primitive data streams: one to receive data from the generating sub-detectors, one to send it 
to the L0TP (section \ref{sec:l0tp}) and one to mirror such data and send it to the acquisition PC; two more ports per switch are used 
to configure the network.

On the acquisition PC, based on Linux\footnote{CentOS 7 64-bit \cite{CentOS}.}, a number of control programs (daemons), managed by the 
experiment Run Control system through DIM, continuously run one instance of the acquisition software for each primitive-producing sub-detector. 
The daemons listen to specific Ethernet ports (one per sub-detector) and acquire MTPs sent by the sub-detectors, performing a fast analysis on 
them and storing them to local disks.
The primitive acquisition software is based on standard C++ Ethernet sockets, and is optimized in terms of performance in order to be able to 
acquire all the MTPs produced in a burst. It uses DIM for synchronization with the experiment, receiving the run and burst number and the SOB 
and EOB signals.

Each primitive-generating sub-detector delivers one MTP per 6.4~$\mu$s during the $\sim$ 5~s long burst; the number of primitives generated 
by sub-detectors in 2016-2018 at $\sim$ 60\% of nominal beam intensity roughly varied from $\sim 5\cdot 10^6$ (LKr) to $\sim 30\cdot 10^6$ 
(NA48-CHOD) per burst, resulting in raw (binary) files of the order of 1 GB per burst. The amount of data produced imposes the use of a 
downscaling factor for permanent storage. Once every $N$ bursts the complete set of generated primitives is acquired, while for all other 
bursts just one MTP every $D$ is acquired; the $N$ and $D$ values are set according to run conditions and beam intensity, and are normally both 
of order $\sim$ 10.
The non-downscaled primitive data files are stored on the CASTOR CERN hierarchical storage system \cite{CASTOR}, from where they can be 
retrieved for trigger efficiency studies or debugging purposes. The reduced files with ``downscaled primitives'' are instead stored temporarily 
on a local disk of the acquisition PC, promptly decoded and used by an on-line monitoring system, which shows information on a display in the 
control room.


The primitive fine-time information was used to synchronise all the sub-detectors participating to the L0 trigger: the time correlation between 
the primitives from RICH and other sub-detectors was used for relative time alignment.
The decoding and analysis program and the correlation algorithm are based on ROOT \cite{root}.
The acquisition program also produces ASCII files containing beam time profile information used for monitoring  and to provide feedback to 
the SPS operators.

\section{The L0 Trigger Processor}
\label{sec:l0tp}

The single L0TP \cite{l0tp} has the task of collecting the L0 trigger primitives from participating sub-detectors, time-aligning them and 
comparing them to a set of programmable conditions (``masks''), in order to generate L0 triggers which must be precisely timed in order to be 
dispatched as time-synchronous pulses. It does so while allowing flexible source masking and trigger down-scaling, and can also autonomously 
generate special L0 triggers for different purposes. 

\subsection{Hardware}
\begin{figure}[hbt] 
  \centering
  \includegraphics[width=1.0\linewidth]{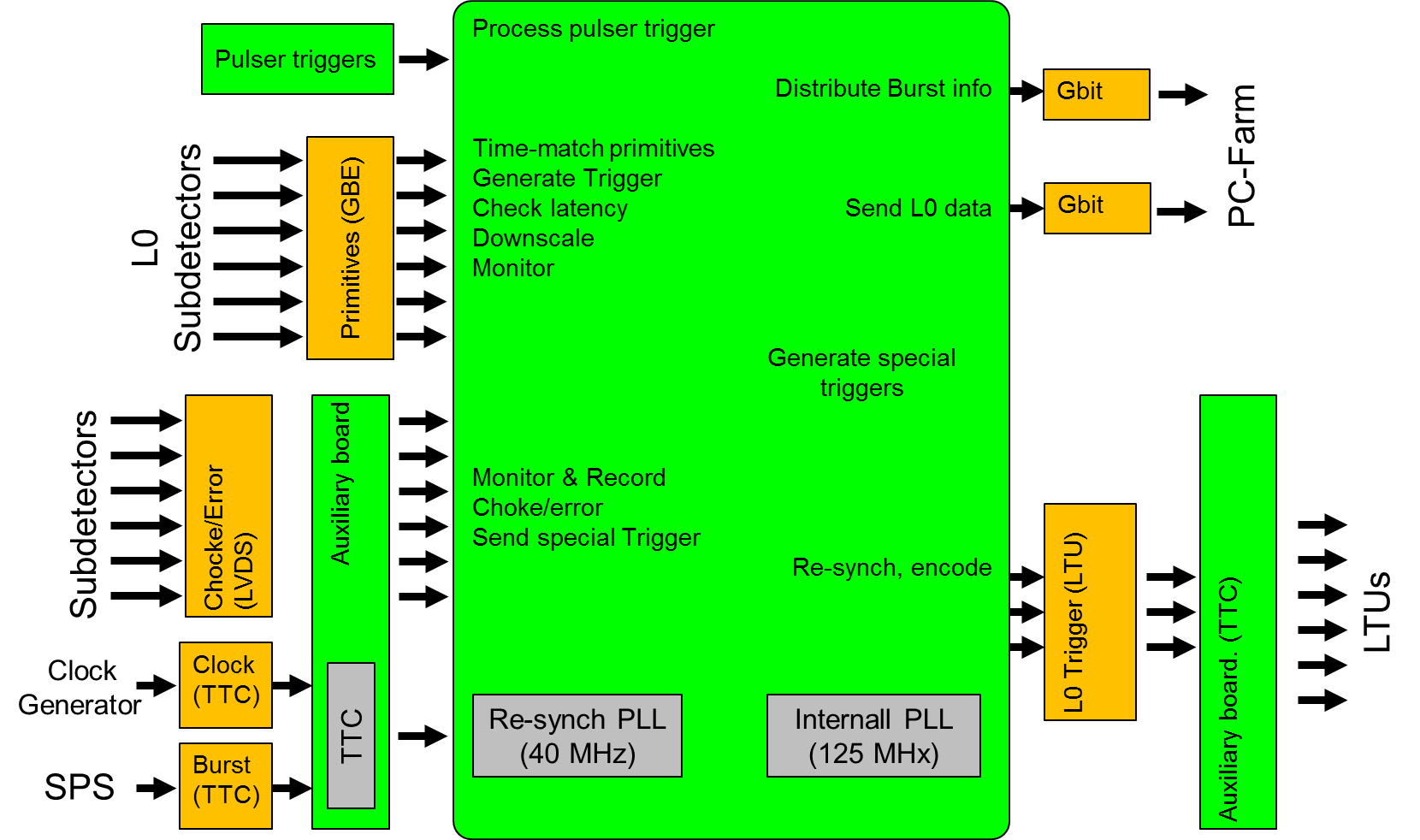}
  \caption{Overview of the logical connections of the L0 Trigger Processor.}
  \label{fig:L0TP_Scheme}
\end{figure}

The L0TP is implemented on a commercial FPGA development board\footnote{Altera DE4 Development and Education board by Terasic Inc.}, hosting 
an Altera Stratix\reg~IV FPGA\footnote{Altera EP4SGX530KH40C2, with 530K logic elements, 27~MB of embedded RAM and highest speed 
grade.}. 
A simplified scheme of the L0TP connections is shown in fig.~\ref{fig:L0TP_Scheme}.

L0 trigger primitives are received from the participating sub-detectors (sources) as MTPs (UDP packets) over GbE.
The development board hosts four GbE ports connected with serial SGMII interfaces to the FPGA; four additional GbE ports were added by using 
two daughter cards with two GbE ports each\footnote{Terasic HSMC-NET.}, connected with a parallel RGMII interfaces, resulting in a total of 
8 GbE ports.
Seven of the GbE links are used for receiving L0 trigger primitives and one is reserved to transmit detailed L0 trigger information to the 
PC farm.

A custom auxiliary daughter card is used to interface the L0TP with the master clock delivered via TTC, through the use of a TTCrq daughter 
card\cite{TTCrq} (hosting an opto-receiver, TTCrx decoder chip and QPLL), and to dispatch the LVDS L0 triggers 
to the first sub-detector LTU 
module (section \ref{sec:common}) of the distribution daisy-chain. The auxiliary card also collects the CHOKE/ERROR signals from each sub-
detector. 

\subsection{Functional overview}
The L0TP firmware consists of the logic to receive MTPs via GbE, a module to time-align them, the logic to time-align the primitives, and 
the algorithm to check trigger matching conditions using an associative memory. Other modules handle the dispatching of L0 trigger 
information to all TDAQ devices and to the PC farm.
The L0TP logic was implemented entirely using the internal FPGA resources, without using any external memory device.

There are two clock domains used in the L0TP firmware: the master 40~MHz clock from the TTC distribution system, and an internally-generated 
125~MHz clock from a PLL fed with the internal 50MHz oscillator of the development board. 
The former is used to synchronize the L0TP with the rest of the experiment and to drive all outputs, while the latter is used for the trigger 
algorithm logic and GbE communication modules. 
Dual-clock FIFO buffers and multiple synchronization flip-flops were used for safely interfacing the two clock domains.

L0 trigger primitives are generated asynchronously by the sub-detectors that participate in the L0 trigger.
The latency of the primitive generation strongly depends on the complexity of the sub-detector and the algorithms used to produce them.
For this reason L0 trigger primitives related to the same physical event in the detector, but generated by different sub-detectors, can be 
delayed with respect to each other.
By imposing a 6.4~$\mu$s time-frame structure on the dispatch of MTPs from the sub-detectors, the delays can be defined in terms of an integer 
number of MTPs; thus primitives related to the same event might be stored in MTP $N_i$ for a faster source and in MTP $N_j > N_i$ for a slower 
one. Such an MTP offset must be compensated by the L0TP. 
Moreover the L0TP must allow enough flexibility in the time-alignment procedure to accommodate fluctuations in the primitive-generation latency 
and the latency associated to the transmission of the primitives through the network.


\subsection{Input and frame alignment}
An UDP/IP packet handling module (ethlink) was developed to send/receive information through the GbE links, interfacing the internal logic with 
the external physical interface devices.
A hardware UDP/IP stack transmits and receives Ethernet frames with a maximum payload length of 1500 bytes; for simplicity, L0TP does not 
support fragmentation or jumbo frames. 
The ethlink module architecture is optimized to sustain the theoretical maximum throughput of the GbE standard without any data loss, managing 
up to 1,448,000 (81,274) frames/s with minimum 46 B (maximum 1.5 kB) frame payload.

An extraction module reads L0 trigger primitives from the ethlink, processing the seven GbE links in parallel: one 32-bit word is extracted 
every 4 clock cycles (32~ns).
For each incoming primitive, the extraction module performs some time consistency checks, to guarantee the correctness of the time frame 
structure and prevent misbehaviour in case primitives arrive too late with respect to the corresponding event time and the maximum allowed 
L0 trigger latency; if inconsistencies are detected the data is discarded and a fatal error flag is raised.

The extracted primitives enter a programmable delay module which compensates for any MTP offsets $\Delta N_i$ between primitive sources, 
in order to realign packet $N$ from the fastest source with packet $N_i = N + \Delta N_i$ from source $i$.
The primitives from the fastest source (\emph{e.g.} those with 0 MTP offset) are stored in dedicated FIFO buffers, while the first 
$\Delta N_i$ frames from slower ones are skipped because these MTPs do not contain viable primitives by definition. 
When the first non-skipped MTP arrives from the slowest source, the delay module reads it directly from the ethlink while reading the first 
packet stored in the buffers for the faster sources.
By continuing to read from the buffers and the ethlink in this way, the MTP offset is taken into account.
The FIFO buffers used to store primitives from the fastest sub-detectors are 8K words deep, thus posing no practical limitation to the maximum 
possible delay at the expected primitive rate (up to 800~$\mu$s for a 10~MHz primitive rate).
During the 2017 run a 3-frame delay (19.2~$\mu$s) was introduced only for the calorimetric L0 primitive generator, but larger values are 
expected when GPU-based primitive generators are introduced.


\subsection{Primitive alignment}
L0 trigger primitives pass from the delay module to a second time-alignment module, in which they are stored in circular RAM buffers, one per 
source, using the primitive time to generate the memory address. Some of the least significant bits of the (25~ns LSb) timestamp and some of 
the most significant bits of the (100~ps LSb) fine-time are used to generate the memory address; the number of fine-time bits, and therefore 
the time ``granularity'' of each memory location, is set by a programmable parameter. At most three fine time bits can be used, so the finest 
possible granularity is 3.125~ns per memory location. This approach results in a rough time alignment of the primitives.
Each memory location can store one primitive, therefore primitives from the same source with time differences smaller than the granularity 
might overwrite each other; the probability of such overwriting depends on the primitive rate, thus ultimately on beam intensity, and can 
be measured using the independent primitive acquisition system (section \ref{sec:primitive_readout}). Such overwriting would result in some 
primitives being lost, and is avoided by ensuring that the sources only provide a single primitive for physical events closer in time than the 
finest L0TP granularity, or even further spaced in time, according to the intrinsic time resolution of sub-detectors.
Each time-alignment buffer has 16K locations, in which both primitive IDs and times are stored; the maximum span for time alignment depends on 
the granularity, and is 51~$\mu$s when using the full 3.125~ns granularity.

Writing into the time-alignment buffers by using the primitive time results in sparse data, and time constraints do not allow a complete scan 
of the (mostly empty) entire buffer for reading at the rate of one location per granularity time unit, required to keep pace with the 
writing. 
To overcome this issue, the L0TP requires one (configurable) source to act as a ``reference sub-detector''. The memory addresses of all reference 
sub-detector primitives are not only stored in the time-alignment buffer, but also in a separate, dedicated FIFO buffer. The latter buffer is 
used to mark the memory addresses that will be considered by the reading process. \\
With the above approach, the reference sub-detector must be included in every L0 trigger mask: this is a limitation, in that it would make 
impossible to measure the trigger efficiency of such sub-detector from the data. 
Thus a second source is required to be a ``control sub-detector'', with primitives from the control detector stored in another dedicated FIFO 
buffer. The control primitives are used to read from the time-alignment buffers, independently from the reference detector but in an identical 
way.
Therefore, the L0TP ultimately generates two types of L0 trigger: one driven by the presence of a trigger primitive from the reference 
sub-detector, and the other from the control sub-detector. The two trigger samples are correlated through their origin from common 
physics events, but are not correlated in terms of sub-detector and L0 trigger primitive generation logic, so the events triggered by the 
control sub-detector can be used to measure the efficiency of the reference sub-detector trigger (and vice versa), and therefore ultimately 
the efficiency of any trigger condition.
During the 2016-2018 data-taking the RICH was typically used as the reference sub-detector, while the NA48-CHOD was mostly used as the 
control sub-detector.

Due to the time constraints in reading the time-alignment buffers, the L0TP cannot clear the time alignment buffers whenever the writing 
address corresponding to input primitives rolls over the buffer address size. In order to distinguish old and new primitives, some of the 
most-significant bits of each primitive time stamp (the lowest changing every 51.2~$\mu$s) are also written in the alignment buffers, and are 
checked at reading time.

\subsection{Primitive matching}
In order to cope with some variability of primitive generation latencies from any source, which could cause primitives related to the same event 
to occur in different MTPs for different sources, the L0TP waits for some programmable time (specified in terms of number of received MTPs) 
before starting to read primitives from the time-alignment buffers.
The reading process starts when a ``received MTP'' counter reaches a given (programmable) value. The read address is taken from the first primitive 
of the reference or control sub-detector, and the contents of all time-alignment buffers are read at that address.
As mentioned, primitives for which the most significant bits of the timestamp do not match those of the current cycle of memory reading are 
ignored.
Primitives related to the same event from different sources may have small time differences, and can therefore be in adjacent memory addresses.
To avoid any trigger inefficiency due to such edge effects, every time a memory location is read, both the previous and the next locations 
are also read.

A time selection around the reference sub-detector time is performed at this time, using the full fine-time information.
Each primitive in the three memory locations that are read is considered a valid match only if it is within a (programmable) time window around 
the reference sub-detector primitive time. In such case it is passed to the Associative Memory module (AM).
The time window sizes can be different for each source, and are of order $\pm$5~ns, reflecting the on-line time resolution of the 
sub-detectors.

The AM consists of two parts: the first is a shift register, which stores the primitives that were read from the time-alignment buffers for 
all sources.
A bitwise OR of the primitive ID of each matching primitive is performed, making an overall primitive ID for each source, with associated trigger 
time information from the reference (or control) detector primitive time. 
Such ID is then managed by the second part of the AM, acting as an associative PROM, comparing the overall primitive IDs of all sources against 
a programmable table of valid L0 trigger masks. Such masks define all the conditions that should generate a L0 trigger. Each bit of a trigger 
mask can be set as required, prohibited (\emph{e.g.} for vetoing sub-detectors), or ``ignore'', and a bitwise AND of all non-ignored bits defines 
a mask as being matched.
The L0TP allows up to 16 independent trigger masks, which are all checked in parallel in one clock cycle. 

To avoid exceeding the maximum L0 trigger rate, triggers from each mask can be down-scaled by a mask-dependent factor $D$, meaning 
that only one out of $D$ occurrences of a mask being matched actually causes a L0 trigger to be produced. The generated L0 trigger-type only 
encodes the masks which were satisfied and also accepted by the down-scaling algorithm. However, all information about matched masks is made 
available in the data through the PC farm.

\subsection{Output stage}
When at least one of the L0 trigger masks is matched and not downscaled, a L0 trigger is produced. It must then be dispatched to all 
sub-detectors after a programmable fixed latency $T_L$ with respect to the originating event.
A 32K-word deep circular buffer, with each location corresponding to 25~ns giving a maximum storage time of 800~$\mu$s, is implemented as a 
dual-port RAM. The buffer is used to synchronise the triggers to their originating events.
The instruction for L0 triggers to be dispatched is stored in the buffer at a location determined by the least significant bits of the 
overall trigger time.

Reading from the buffer is delayed by the ``latency time''. At the beginning of the burst a (25~ns period) timestamp counter starts counting, 
while the buffer read-address remains idle. When the counter reaches a programmed value, corresponding to the overall L0 trigger latency time, 
the read-address starts incrementing from the first buffer location with the same 25~ns period, thus reading location $N$ at the time 
$N+T_L$/25~ns, while at the same time clearing the location.
During the 2016-2018 run the actual latency used was 197~$\mu$s, but larger values will be required when introducing GPU-based primitives 
(section \ref{sec:gpu}).

If the read location contains trigger information matching the current read cycle, such information is used to dispatch a L0 trigger to 
all sub-detectors. The L0 trigger consists of a 25~ns long synchronous pulse giving the trigger time information, and an asynchronous 
6-bit trigger-type word which can be used by sub-detectors for selective readout. 
Detailed information is dispatched to the PC farm as the L0TPs event data, which allows off-line reconstruction of features of the triggered 
event. 
Such data includes the fine-time and ID of the primitives in all three buffer locations from each sub-detector, the event timestamp, and a 
``trigger flag'' that encodes which of the trigger masks were matched.

The L0TP can generate so-called special triggers autonomously. The Start-of-Burst (SOB) trigger is the first trigger dispatched by the L0TP at 
the beginning of each burst, and the End-of-Burst (EOB) trigger is the last trigger dispatched at the end of each burst. 
The EOB trigger is used to request data frames containing summary, statistics and monitoring information from each sub-detector, which are 
collected together with the main data; one mandatory piece of information produced by each sub-detector is the local timestamp value at 
which the EOB signal was received: this allows an equality check on the number of clock cycles counted by each sub-detector during the burst, 
thus ensuring consistency of time measurements.

Besides trigger primitives, the L0TP also handles asynchronous CHOKE/ERROR signals from the sub-detectors, which are used to monitor overload 
conditions in the readout systems. Whenever one of such signals is active from any (unmasked) source, no L0 triggers are dispatched by the L0TP. 
In order to allow data acquisition efficiency and integrity to be assessed at all times, both the start and the end of any CHOKE/ERROR condition 
are signalled to all sub-detectors and the PC farm by delivering special triggers, which must be acknowledged as usual.
 
To avoid exceeding the design L0 trigger rate of 1~MHz, an \emph{autochoke} protection mechanism was implemented. The autochoke causes the L0TP 
to stop dispatching triggers (notifying all systems) in case beam intensity fluctuations resulted in instantaneous rates above such value.
When a programmable ``full'' level is reached in the output-link buffers, the L0TP continues to process triggers but does not dispatch them 
(except for the appropriate CHOKE special trigger).

Random, periodic and calibration triggers are also generated autonomously by the L0TP for testing and monitoring purposes.
Random triggers are generated using linear feedback shift registers, with a trigger being produced when the least-significant bit of a 
pseudo-random number is set. The rate of random triggers is a programmable parameter of the L0TP. These triggers are used to test the 
performance of the TDAQ system. 
A pulse generator is used to produce two different periodic trigger flows, with independently programmable periods in units of 25~ns. 
Such triggers are used to measure noisy channels in the GTK and to monitor the pedestal values of the cells of the LKr electro-magnetic 
calorimeter. 
Both random and periodic triggers can be inhibited during part of the burst in a programmable way.

Finally, several sub-detectors need special triggers for calibration purposes: when a sub-detector participating to L0 trigger generation 
sends primitives with the most significant bit set, the L0TP forces the generation of a trigger, skipping the alignment and time-matching 
logic entirely.




%
%

\section{GPU-based trigger}
\label{sec:gpu}

The relatively large latency of the NA62 L0 trigger allows the use of selection criteria usually applied at higher trigger levels.
Although the processing power of FPGA devices is steadily increasing, non-reconfigurable processors (CPUs or GPUs), are still unbeatable for 
problems requiring complex calculations. 

As a pilot project, we studied the possibility of using Graphics Processing Units (GPUs) to generate complex trigger primitives at the lowest 
trigger level in NA62.
The parallel architecture of GPUs can be exploited to parallelize complex reconstruction algorithms or to simultaneously process several 
independent and uncorrelated events. 
In contrast to more common computational uses of GPUs, their use in a hard real-time scenario requires consideration of the total latency 
besides the sheer computing throughput. 
Because of their many-core architecture, GPUs reach their theoretical peak performance only when processing large enough datasets. 
This results in an increase of the latency, which can be an issue in a system where the arrival times of the events to be processed fluctuates 
randomly. Moreover, the total latency is also affected by the transfer time of events from the readout system to the GPU memory and the time 
required to initiate processing and deliver results to their consumer.

The NA62 GPU trigger system consists of four parts: TEL62 logic, Network Card, GPU kernel and Synchronizer. 
The data are pre-processed in TEL62 boards and then sent to the GPU through a custom reconfigurable PCI Express (PCIe) Network Interface Card 
(NIC), called NaNet~\cite{Nanet}. NaNet allows real-time processing, by handling the data streams coming from the readout boards and sending 
the reconstructed events directly to the GPU memory, coordinating with the host CPU to launch the processing stage.
The events, processed by a dedicated GPU kernel, are then sent to a Synchronizer to adapt the GPU results to the periodic primitive structure 
required by the L0TP for the matching with other sub-detector trigger primitives. \\
As a first application, we focused on the possibility to reconstruct the \v{C}erenkov rings produced by charged particles in the NA62 RICH
detector (GPURICH trigger). 

\subsection{TEL62 GPURICH logic}
While the standard L0 trigger logic for the RICH (section \ref{sec:RICH}) produces multiplicity-based primitives working on 8-channel analogue 
sums, the 4 TEL62s used for readout also produce a flux of compressed data for GPU elaboration, containing information on all the PM hits.

\begin{figure}[htb]
  \centering
  \includegraphics[width=0.9\linewidth]{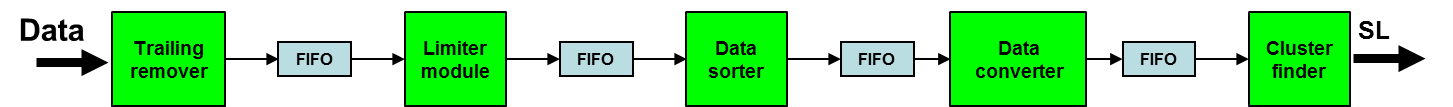}
  \caption{Block diagram of the GPURICH logic in the PP-FPGA.}
  \label{fig:GPU_PP}
\end{figure}

Fig.~\ref{fig:GPU_PP} illustrates the GPURICH logic in the TEL62 PP-FPGA. 
The data reach the GPURICH module through the common interface, and in the Trailing Remover module only the leading-edge time measurements are 
kept. Channel time offsets are also added by this module. Since hits in the TDCB 6.4~$\mu$s frames are not time ordered, a sorting network 
with a single pipeline and a fixed number of sorting cells is implemented in the Data Sorter module (DS). The latency of this module is kept 
constant using a delay stage. 
The number of sorting cells can be chosen depending on the maximum number of hits in a frame at full beam intensity, currently being 270.

Since beam extraction instabilities can lead to rate fluctuations exceeding limits, a Limiter module (LI) at the input of the DS limits the 
number of hits to the allowed maximum. In the Data Converter module (DC) the format is changed to set the absolute hit times in 40-bit words 
before a Cluster Finder (CF), the last logic module, in which time-ordered hits are grouped in clusters according to a programmable time 
window starting at the first hit. For each cluster the average time is computed, and the channel identifier is encoded into 9 bits.

\begin{figure}[htb]
  \centering
  \includegraphics[width=0.9\linewidth]{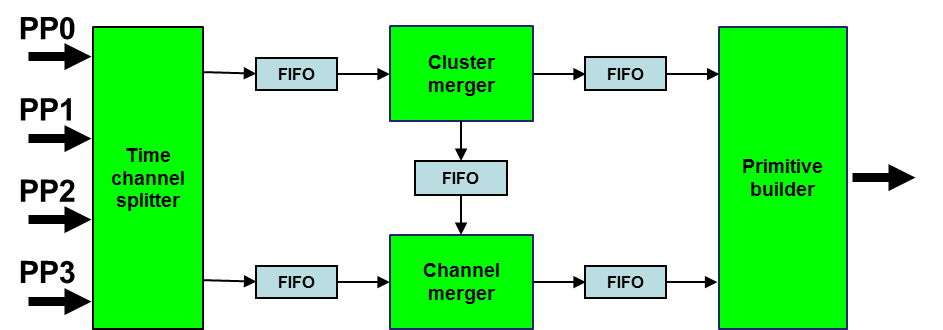}
  \caption{Block diagram of the GPURICH logic in the SL-FPGA.}
  \label{fig:GPU_SL}
\end{figure}

Data from the 4 PP-FPGAs are merged in the SL-FPGA, whose logic is shown schematically in fig.~\ref{fig:GPU_SL}.

The Time-Channel Splitter module (TCS) waits for data from the four sources and splits time and channel information in two parallel branches. 
In the time branch a clustering algorithm similar to the one described above is applied by the Cluster Merger module (CLM), also computing
the average time. Channel information is requested by the CLM to the Channel Merger module (CHM) for each defined cluster. 
The complete event is built in the Primitive Builder module (PB). Multiple events are stored into UDP packets called MGPs (Multi GPU Packets), 
which are sent via GbE every 12.8~$\mu$s. 

The total produced event rate at full beam intensity is about 15~MHz, with an average event size of 28 bytes, resulting in a $\approx$ 120~MB/s 
data rate from each TEL62, requiring the use of two of the four GbE links in each TEL62 (a total of 8 links) to feed the GPU trigger system.
The total latency of the GPURICH firmware is $\sim$ 2.5~$\mu$s, with RMS fluctuations well below the 6.4~$\mu$s frame duration.

\subsection{NaNet}
\label{par:nanet}
NaNet is a \mbox{FPGA-based} reconfigurable PCIe NIC implementing direct \mbox{zero-copy} communication between network channels and the memory 
of the host CPUs and NVIDIA GPUs (GPUDirect\reg \, RDMA).
The design is modular and supports different network link technologies, allowing deployment in multiple scenarios: standard GbE 
(\mbox{1000BASE-T}) and 10 GbE (\mbox{10GBase-KR}) plus custom 34~Gbps APElink~\cite{apenet} and 2.5~Gbps deterministic latency 
KM3link~\cite{nanetNSS}. 
The management of the supported network protocols (\emph{e.g.} IP and UDP) is performed in the FPGA logic in order to avoid OS-related time 
jitter and to guarantee a deterministic communication latency while achieving maximum capability for the channel. A high-performance 
PCIe Gen2/3 DMA engine performs \mbox{zero-copy} data transfers to and from application CPU and GPU memories.
The design also implements a reconfigurable processing stage on both inward and outward data streams, enabling the implementation of 
heterogeneous stream-processing pipelines having CPUs, GPUs and FPGAs as building blocks.

The measured total time latency of these heterogeneous processing pipelines and their stability have been assessed in several experimental 
contexts~\cite{NanetVLVnT}\cite{nanet_development}.
Data received from TEL62 boards are collected in a Circular List Of Persistent (CLOP) buffers in GPU memory. The size of each CLOP is defined 
by a configurable time-out (``gathering time''); when a CLOP buffer is ready an ``RX done'' event is also \mbox{DMA-written} by NaNet in a host 
CPU memory region (event queue) and trapped in \mbox{kernel-space} by a device driver, which sends a signal to the user application, which in 
turn launches the kernel on the GPU.

Once the kernel has completed, results can be sent via NaNet to a remote device.
Data are \mbox{DMA-read} from GPU memory; the kernel device driver instructs the NIC by filling a ``descriptor'' into a dedicated
\mbox{DMA-accessible} memory region (``TX ring'').
The presence of a new descriptor is notified over PCIe to NaNet by writing on a doorbell register so that the board can issue a ``TX done'' 
event in the event queue.
  


\subsection{GPU Kernel}
Data are processed as soon as they are copied into GPU memory.
As in any real-time environment, algorithms must be fast enough to cope with the input event rate while remaining within the L0TP latency. 
Moreover, as data from other sub-detectors are not available at this trigger stage, pattern-recognition algorithms need to be 
\emph{seedless}, \emph{i.e.} not relying on any externally-provided ring centre information.

We focused on two multi-ring pattern recognition algorithms based only on geometry. 
In the first one (``Histogram'' \cite{acat2017}), the plane on which the RICH PMs lie is divided into a grid and a histogram is created with 
distances between grid points and the coordinates of the centre of each hit PM. \v{C}erenkov rings are identified by bins whose contents exceed 
a programmable threshold. 
The second algorithm (``Almagest'' \cite{almagest}) is based on Ptolemy's Theorem.
Both algorithms are used for pattern recognition, and once the number of rings and the points belonging to each of them have been identified 
Crawford's method \cite{Crawford} is used to obtain center coordinates and radii with improved resolution.

\subsection{Synchronizer}

Due to parallel computing, the GPU output is asynchronous to the event time and not time ordered in each GLOP (a GLOP being a GPU-processed CLOP).
In order to cope with the periodic 6.4~$\mu$s MTP structure required by the L0TP (section \ref{sec:l0tp}), a synchronization stage is 
implemented, for the time being within an additional FPGA development board\footnote{Terasic DE4, equipped with an Altera 
Stratix\reg~IV FPGA.}, to be later migrated inside NaNet.

A GLOP (comprising 256~$\mu$s of data), is split in 40 miniGLOPs (6.4~$\mu$s each) based on the event timestamp information. 
MiniGLOPs are read sequentially to preserve the time structure, with two miniGLOP buffers allowing concurrent read and write operations.
The latency of this stage is 512~$\mu$s, for which the L0TP can compensate by using digital delays (see section \ref{sec:l0tp}).
   
\subsection{Integration}
The whole system was installed close to the RICH readout electronics rack in the NA62 experimental hall, and was run parasitic to the standard 
trigger during the 2017-2018 data-taking period. For this experimental set-up we tailored the NaNet design to be implemented on an FPGA 
development board\footnote{Terasic DE5-net, hosting an Altera 5SGXEA7N2F45C2 FPGA.} equipped with an Altera Stratix\reg~V FPGA with four 
SFP+ connectors hosting 10GbE channels and a PCIe Gen2 x8 bus connection with the host PC (NaNet-10 \cite{nanet10}).

A test-bed consisted of a commercial Ethernet switch\footnote{Hewlett-Packard HP2920.}, a NaNet-10 PCIe board plugged into a PC 
host\footnote{\mbox{X9DRG-QF} \mbox{dual-socket} motherboard with Intel Xeon\reg \mbox{E5-2620} 2 GHz CPUs (Ivy Bridge), 32~GB of 
DDR3 RAM.} together with an NVIDIA Pascal P100 GPU. 

The processing pipeline is inset between the RICH readout and the L0TP. Data from the detector PMs are collected by four TEL62 readout boards 
sending primitives to NaNet-10 as UDP datagram streams over two GbE channels connected to the GbE/10GbE network switch. Packets are then routed 
on a 10GbE channel towards one of the NaNet-10 ports. A processing stage on the on-board FPGA decompresses and coalesces the event fragments 
scattered among the eight UDP streams according to a configurable time-window; a zero-copy DMA transfer towards the GPU memory is then 
instantiated to transmit the reconstructed events to the GPU memory. 

\begin{figure}[!hbt]
  \centering
  \includegraphics[width=\linewidth]{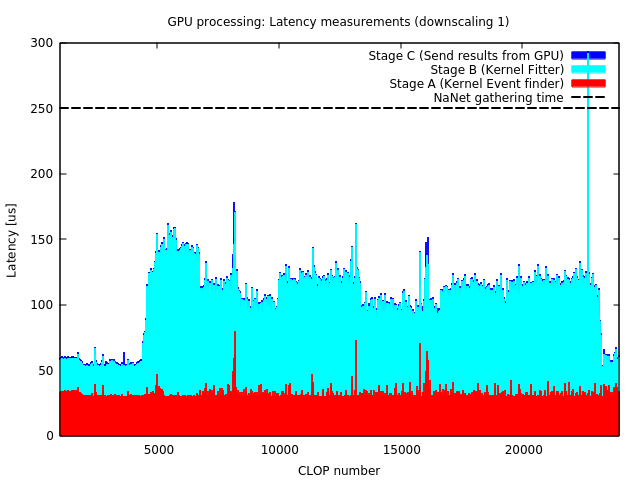}
  \caption{GPU system heterogeneous processing pipeline time latency. The lower histogram (red/grey) indicates the (almost constant)   
  latency of event indexing in the CLOP buffer, the upper one (cyan/light grey) the histogram-based ring reconstruction kernel latency, 
  and the barely visible topmost one (blue/dark grey) that for the transmission of results. 
  Data refer to a single burst at beam intensity $19 \cdot 10^{11}$ protons per burst, with a NaNet gathering time of 250~$\mu$s 
  (dashed line).}
  \label{fig:GPU_latency}
\end{figure}

Latencies related to GPU processing (event indexing and ring reconstruction) and transmission of results to the L0TP (via UDP) are shown
in fig.~\ref{fig:GPU_latency}: the overall time is always well below the maximum allowed L0 latency. 
The reconstructed ring resolution is comparable to that obtained by off-line algorithms, thus proving that the system can be used as conceived.

\section{Running experience}
\label{sec:running}

The NA62 L0 trigger system was tested during the 2015 and 2016 data-taking periods without the calorimetric part, and has been fully operational 
since the beginning of the 2017 data-taking period.
The beam intensity during 2017 was of order $18 \times 10^{11}$ protons on target during a burst, corresponding to about 55\% of nominal. 
The corresponding hit rates per channel during the burst varied among sub-detectors, from about 1.5~MHz per TDCB for LAV to about 
3.6~MHz per TDCB for KTAG and CHANTI (not involved in the L0 trigger).

\begin{figure}[hbt]
  \centering
  \includegraphics[width=0.9\linewidth]{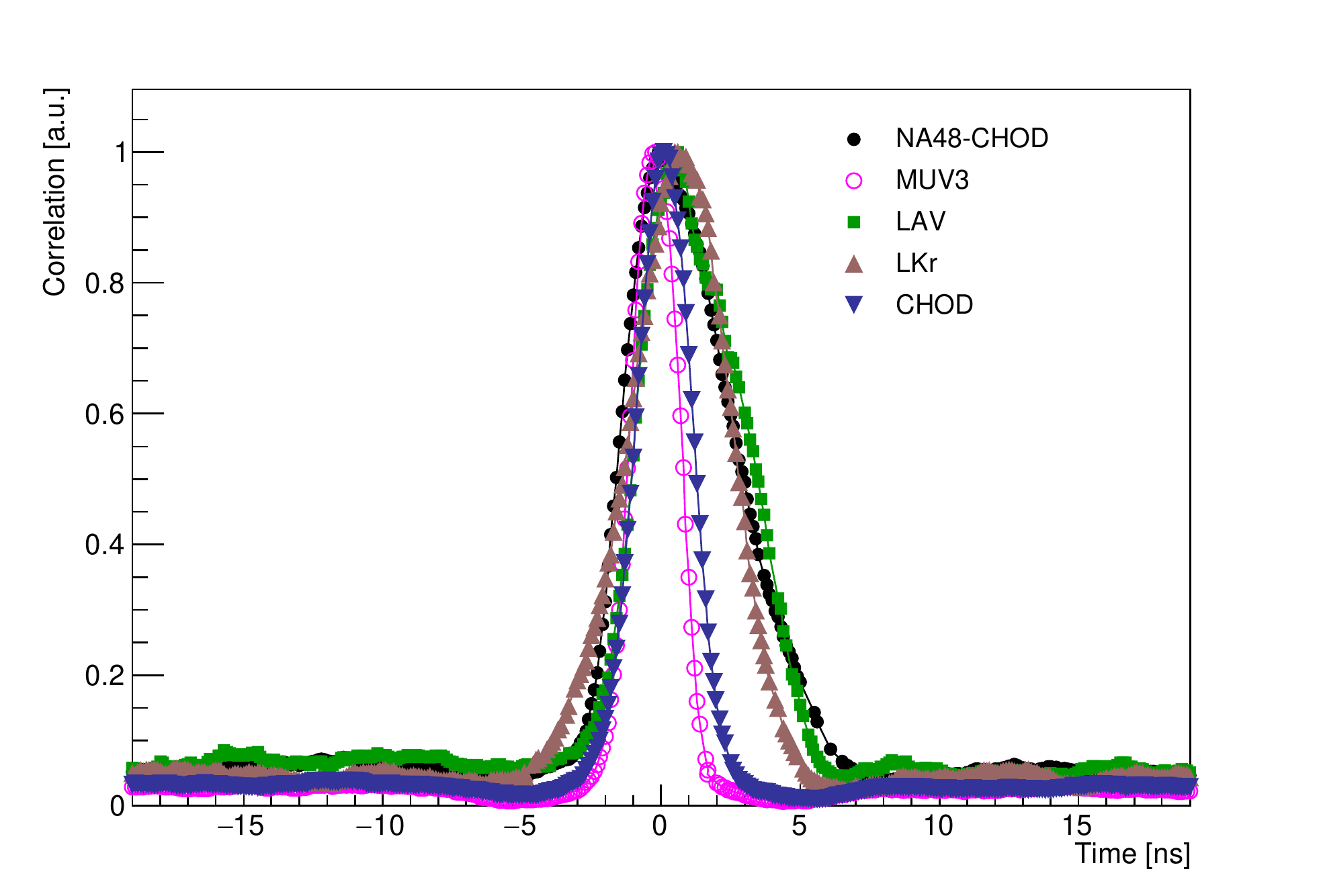}
  \caption{L0 trigger primitive time correlation between RICH and other sub-detectors after alignment.}
  \label{fig:primitive_correlation}
\end{figure}

Precise relative time alignment between sub-detectors is essential in order not to compromise the L0 trigger efficiency due to time-matching 
failures, and to allow tighter timing cuts, thus reducing trigger losses due to accidental hits in vetoing sub-detectors. Such alignment is 
achieved by analysing the time-correlation of the L0 trigger primitives, shown in fig.~\ref{fig:primitive_correlation}, as monitored by the 
independent primitive acquisition system (section \ref{sec:primitive_readout}).

The average number of generated L0 trigger primitives per burst were: $\approx 8 \cdot 10^6$ from the calorimetric system, 
$\approx 30 \cdot 10^6$ for RICH, and $\approx 34 \cdot 10^6$ for MUV3, CHOD and NA48-CHOD. Considering a burst duration of $\sim$ 3.5~s and 
the simultaneous activation of different trigger masks, the average instantaneous L0 trigger primitive rates match the design value of order 
10~MHz from TDC-based systems, but this value turned out to be quite variable during the burst.

\begin{figure}[hbt]
  \centering
  \includegraphics[width=0.9\linewidth]{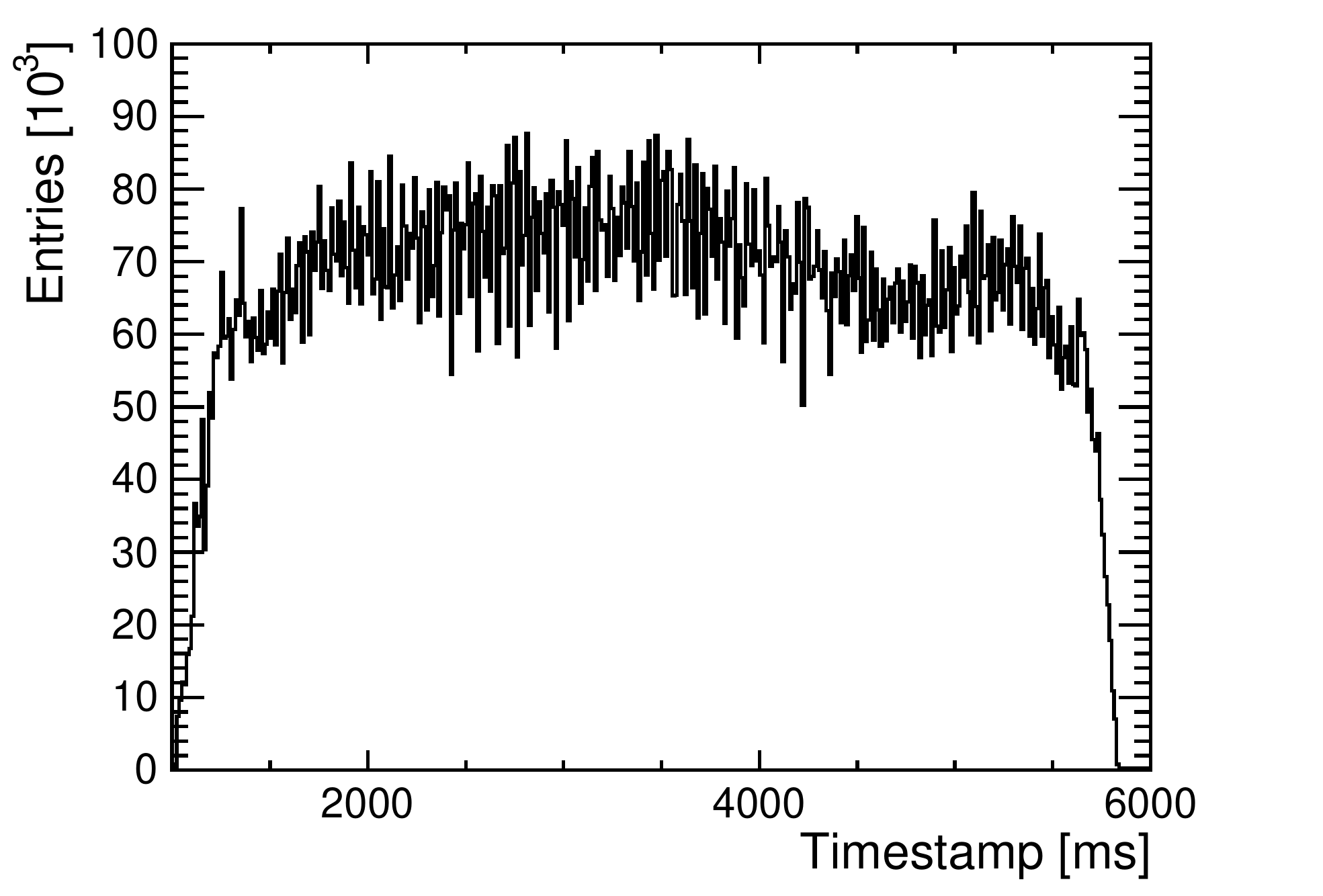}
  \caption{Beam intensity profile along a burst (in ``good'' conditions) as determined by the NA48-CHOD L0 trigger primitive standalone readout;    
  each time bin is 20~ms wide.}
  \label{fig:beam_profile}
\end{figure}

The independent acquisition system for primitives (section \ref{sec:primitive_readout}) allowed, through their timestamp values,  
an unbiased monitoring of the instantaneous beam intensity during the burst, which exhibits quite large fluctuations, as shown in 
fig.~\ref{fig:beam_profile}. Depending on the quality of the beam extraction, the beam intensity profile can have even more dire fluctuations, 
resulting in instantaneous primitive rates up to 16~MHz, with a peak-to-valley ratio up to a factor 3.

While the TDAQ system has de-randomizing buffers at all stages in order to accommodate random rate fluctuations, the time scales for such 
beam intensity peaks extends to tens of ms, \emph{i.e.} effectively infinite with respect to the time scale of the logic, periodically 
exposing the system to an effectively constant rate quite above the design value, which can result in choking and data loss in several 
places. When extreme beam fluctuations persisted for a few burst, improvements on beam extraction were requested to the SPS control room, 
to guarantee acceptable beam intensity profile and ensure stable data-taking conditions.

About $1.8\cdot 10^{6}$ L0 triggers were generated in each burst. 
The corresponding data readout bandwidth varied among sub-detectors depending on the hit rate and the number of 25~ns time slots around the 
trigger time being read. In the KTAG (7 time slots per event) this was $\approx$ 72~MB/s, while in the LAV (16 time slots per event) it was 
$\approx$ 44~MB/s. 

Since most of the trigger and readout electronics is in the experimental hall, the environmental radiation can lead to Single-Event Upsets 
(SEUs) and failure in the (non-radiation resistant) FPGA firmware, resulting in system errors.
Logic corruption in the TDCB or PP-FPGA could result in missing regions in the readout channels' occupancy or mismatches between L0 trigger 
and event times; corruption in the SL-FPGA could result in badly-formatted or missing events. 
Missing data packets are easily detected by the Run Control system, since they result in event building failure in the PC farm, and 
an automatic system, continuously monitoring data quality plots, alerts the experiment shift crew about the need to reload the FPGA firmware.
These kinds of errors were found to occur at a rate varying from roughly one per day (for LAV) to roughly one per hour (for KTAG and CHANTI),  
broadly consistent with estimates performed at the design stage.

\section{Summary and perspectives}
\label{sec:conclusions}

A versatile, fully-digital, high-density integrated readout and L0 trigger system was developed for the NA62 experiment. It handles both TDC 
and ADC data from several different sub-detectors by using a powerful highly-customizable carrier board as the main underlying hardware.
The full integration of readout and L0 trigger allows the generation of trigger primitives for the lowest trigger level to process 
the full-granularity data available from sub-detectors. Besides posing no \emph{a priori} limitations on trigger processing, such approach 
presents benefits in terms of economy of hardware, firmware, and an unrestricted monitoring capability. Slow-control information gathering 
was naturally integrated into the system by exploiting the L0 trigger distribution network for commands to the entire system.
A rather large (by HEP standards) L0 trigger latency allows for heterogeneous processing elements, such as the addition of GPUs used in 
hard real time.
The system has been successfully deployed and used in the data-taking phase of NA62; no bottlenecks were identified at all tested beam 
intensities.

While some improvements are being considered in order to increase the data throughput and reduce the sensitivity to environmental radiation, 
the flexibility of the system, which does not preclude in principle any kind of on-line trigger selection, implies that the L0 trigger 
selectivity can be tightened to a large extent by firmware algorithm changes, in an iterative process which proceeds in steps with careful 
analysis of the collected data.
Detailed performance results will be discussed in a forthcoming paper.

\section{Acknowledgements}
The authors are grateful to the whole NA62 Collaboration for its support during the commissioning of the TDAQ system and for its dedication in 
operating the experiment during the data-taking periods.  Many colleagues deserve to be thanked for their contributions and useful discussions on 
the work presented here.
We are particularly thankful to: C.~Avanzini, A.~Antonelli, M.~Bizzarri, N.~De Simone, E.~Imbergamo, R.~Lietava, G.~Magazzu, M.~Moulson, 
M.~Piccini, A.~Salamon, C.~Santoni and  T.~Spadaro.

The cost of the hardware systems described here was supported by the funding agencies of the Collaboration Institutes. 




\begin{thebibliography}{00}

\bibitem{NA62}
The NA62 collaboration, The beam and detector of the NA62 experiment at CERN, JINST {\bf 12} (2017) P05025.

\bibitem{Theory}
J.~Buras \emph{et al.}, $K^+ \rightarrow \pi^+ \nu \overline{\nu}$ and $K_L \rightarrow \pi^0 \nu \overline{\nu}$ in the Standard Model: 
Status and Perspectives, JHEP {\bf 1511} (2015) 033.

\bibitem{BNL}
A.~V.~Artamonov \emph{et al.} (BNL-E949 Collaboration), Study of the decay $K^+ \rightarrow \pi^+ \nu \overline{\nu}$ in the momentum region 
140 $\le P(\pi) \le 199$ MeV/$c$, Phys. Rev. D  {\bf 79} (2009) 092004.

\bibitem{TTC}
B.~G.~Taylor, Timing distribution at the LHC, 8th Workshop on Electronics for LHC Experiments, Colmar, CERN-LHCC-G-014, (2002).

\bibitem{QPLL}
P.~Moreira, QPLL: a Quartz Crystal Based PLL for Jitter Filtering Applications in LHC, 9th Workshop on Electronics for LHC Experiments, 
Amsterdam, CERN-LHCC-2003-055 (2003).

\bibitem{LTU}
A.~Jusko \emph{et al.}, Proc 10th Workshop on Electronics for LHC and Future Experiments, Boston, MA, (U.S.A.), September 2004. 
CERN 2004-010, p. 277. 

\bibitem{TTCex}
B.~G.~Taylor, TTC laser transmitter user manual, available at \path{http://ttc.web.cern.ch/TTC/} (accessed on December 2018).

\bibitem{TTCrx}
J.~Christiansen \emph{et al.}, Receiver ASIC for Timing, Trigger and Control distribution in LHC experiments, IEEE Trans. Nucl. Sci. 43 
(1996), 1773.

\bibitem{DIM}
C.~Gaspar, \emph{et al.}, DIM, a portable, light weight package for information publishing, data transfer and inter-process communication, 
Comput. Phys. Commun. {\bf 140} (2001) 102.

\bibitem{TELL1}
G.~Haefeli \emph{et al.}, The LHCb DAQ interface board TELL1, Nucl. Instrum. Meth. Phys. Res. A {\bf 560} (2006), 494-502.

\bibitem{TEL62}
B.~Angelucci \emph{et al.}, TEL62: an integrated trigger and data acquisition board, IEEE Nuclear Science Symposium and Medical Imaging 
Conference (NSS/MIC), Valencia (2011) 823-826.

\bibitem{QuadGbE}
H.~Muller \emph{et al.}, Quad Gigabit Ethernet plug-in card, LHCb Technical Note (2005), 
\path{https://edms.cern.ch/ui/file/520885/5/tech\_note\_quadgigabit\_ver2.3.pdf}  (accessed on December 2018).

\bibitem{ScientificLinux}
\path{https://www.scientificlinux.org/} (accessed on December 2018).

\bibitem{HPTDC}
J.~Christiansen, High Performance Time to Digital Converter, CERN/EP-MIC, March 2004, 
\path{http://tdc.web.cern.ch/TDC/hptdc/docs/hptdc manual ver2.2.pdf}  (accessed on December 2018).

\bibitem{TDCB}
E.~Pedreschi \emph{et al.}, A high-resolution TDC-based board for a fully digital trigger and data acquisition system in the NA62 experiment 
at CERN - IEEE Trans. Nucl. Sci. 62,3 (2015) 1050.

\bibitem{LAV-fw}
F.~Gonnella, \emph{et al.}, The NA62 LAV front-end electronics and the L0 trigger generating firmware, Proceedings of Science TIPP {\bf 2014} 
(2014) 397.

\bibitem{LAV-frontend}
A.~Antonelli \emph{et al.}, Performance of the NA62 LAV front-end electronics, JINST {\bf 8} (2013) C01020.

\bibitem{TTCrq}
P.~Moreira, TTCrq user manual, CERN-EP/MIC, Geneva (2004), at \path{http://proj-qpll.web.cern.ch/proj-qpll/images/manualTTCrq.pdf}  (accessed 
on December 2018).

\bibitem{Catapult}
Catapult\reg \, high-level synthesis: \path{https://www.mentor.com/hls-lp/catapult-high-level-synthesis}.

\bibitem{CentOS}
\path{https://www.centos.org/} (accessed on December 2018).

\bibitem{CASTOR}
CASTOR project page at CERN: \path{http://castor.web.cern.ch} (accessed on December 2018).

\bibitem{root} 
ROOT Data Analysis Framework: \path{https://root.cern.ch} (accessed on December 2018).

\bibitem{RICH_ref}
M.~Barbanera, F.~Gonnella, Real-time FPGA design for the L0-trigger of the RICH detector of the NA62 experiment at CERN SPS, Journal of 
Instrumentation {\bf 12} (2017) C01023.

\bibitem{intertelMB}
M.~Barbanera, Design and FPGA implementation of the test equipment for a Digital Communication System of the NA62 High-Energy Physics 
experimental platform at the CERN SPS, Master's Thesis, University of Perugia (2015), available at  
\path{http://na48.web.cern.ch/NA48/Welcome/thesis/mthesis_barbanera.pdf} (accessed on December 2018).

\bibitem{intertelML}
M.~Lupi, Development, Implementation and Experimental Assessment of the Digital Communication System for the Level 0 Trigger of the NA62 
Experiment at CERN-SPS, Master's Thesis, University of Perugia (2015), available at  
\path{http://na48.web.cern.ch/NA48/Welcome/thesis/mthesis_lupi.pdf} (accessed on December 2018).


\bibitem{l0tp}
D.~Soldi, S.~Chiozzi, Level Zero Trigger Processor for the NA62 experiment, Journal of Instrumentation {\bf 13} (2018) P05004.

\bibitem{Nanet}
A.~Lonardo \emph{et al.}, {\it NaNet: a Configurable NIC Bridging the Gap Between HPC and Real-time HEP GPU Computing}, Journal of
Instrumentation {\bf 10} (2015) C04011.

\bibitem{apenet}
R.~Ammendola \emph{et al.}, APEnet+ 34 Gbps Data Transmission System and Custom Transmission Logic, Journal of Instrumentation {\bf 8} 
(2013) C12022.

\bibitem{nanetNSS}
A.~Aloisio \emph{et al.}, The NEMO experiment data acquisition and timing distribution systems, Proc. Nuclear Science Symposium and 
Medical Imaging Conference (NSS/MIC) 2011 (2011), 147.

\bibitem{NanetVLVnT}
R.~Ammendola \emph{et al.}, Nanet3: The on-shore readout and slow-control board for the KM3NeT-Italia underwater neutrino telescope, 
EPJ Web of Conferences {\bf 116} (2016) 05008.

\bibitem{nanet_development}
R.~Ammendola \emph{et al.}, Development of network interface cards for TRIDAQ systems with the NaNet framework, Journal of Instrumentation 
{\bf 12} (2017) C03037.

\bibitem{acat2017}
R.~Ammendola \emph{et al.}, \mbox{Real-time} heterogeneous stream processing with NaNet in the NA62 experiment, 
Journal of Instrumentation \textbf{9} (2014) C02023.

\bibitem{almagest}
G.~Lamanna, Almagest, a new trackless ring finding algorithm, Nucl. Instrum. Meth. Phys. Res. A {\bf 766} (2014) 241.
  
\bibitem{Crawford}
J.~Crawford, A non-iterative method for fitting circular arcs to measured points, Nucl. Instr. Meth. Phys. Res. {\bf 211} (1983) 223.

\bibitem{nanet10}
R.~Ammendola \emph{et al.}, Nanet-10: a 10GbE network interface card for the GPU-based low-level trigger of the NA62 RICH detector, 
Journal of Instrumentation {\bf 11} (2016) C03030.

\end{thebibliography}


\end{document}